\documentclass[11pt]{article}
  
\usepackage{amsmath,amssymb}

\usepackage{epsfig}  
\usepackage{graphicx}               
\usepackage{pstricks}
\usepackage{color}

\newcommand{\nc}{\newcommand}  

\nc{\beq}{\begin{equation}}  
\nc{\eeq}{\end{equation}}  
\nc{\beqa}{\begin{eqnarray}}  
\nc{\eeqa}{\end{eqnarray}}  
\nc{\bea}{\begin{eqnarray}}  
\nc{\eea}{\end{eqnarray}}  
\nc{\ra}{\rightarrow}  
\nc{\lsim}{\begin{array}{c}\,\sim\vspace{-21pt}\\< \end{array}}  
\nc{\gsim}{\begin{array}{c}\sim\vspace{-21pt}\\> \end{array}}  
\nc{\Tr}{{\rm Tr}}
\nc{\slsh}{\slash\hspace*{-0.22cm}}

\def\be{\begin{equation}}
\def\ee{\end{equation}}
\def\bea{\begin{eqnarray}}
\def\eea{\end{eqnarray}}
\def\bit{\begin{itemize}}
\def\eit{\end{itemize}}

\newcommand{\missET}{\slash{\hspace{-2.5mm}E}_T}

\def\to{\rightarrow}

\title{  
\vspace*{-2.3cm}  
\begin{flushright}  
\normalsize{  
SLAC-PUB-14810
  }  
\end{flushright}  
\vspace{1.5cm}  
\Large  
\textbf{
 Hiding a Heavy Higgs Boson at the 7 TeV LHC
}\vspace*{1.0cm}   
}

\author{Yang Bai$^{a}$, JiJi Fan$^{b}$ and JoAnne L. Hewett$^{a}$
\vspace{5mm}
\\
$^{a}$ \normalsize\emph{SLAC National Accelerator Laboratory, 2575 Sand Hill Road, Menlo Park, CA 94025, USA} \\
$^{b}$ \normalsize\emph{Department of Physics, Princeton University, Princeton, NJ, 08540, USA}
}

\begin{document}  
\setcounter{page}{0}  
\maketitle  

\vspace*{1cm}  
\begin{abstract} 
A heavy Standard Model Higgs boson is not only disfavored by electroweak precision observables but is also excluded by direct searches at the 7 TeV LHC for a wide range of masses. Here, we examine scenarios where a heavy Higgs boson can be made consistent with both the indirect constraints and the direct null searches by adding only one new particle beyond the Standard Model. This new particle should be a weak multiplet in order to have additional contributions to the oblique parameters. If it is a color singlet, we find that a heavy Higgs with an intermediate mass of 200 - 300 GeV can decay into the new states,  suppressing the branching ratios for the standard model modes, and thus hiding a heavy Higgs at the LHC. If the new particle is also charged under QCD, the Higgs production cross section from gluon fusion can be reduced significantly due to the new colored particle one-loop contribution. Current collider constraints on the new particles allow for viable parameter space to exist in order to hide a heavy Higgs boson. We categorize the general signatures of these new particles, identify favored regions of their parameter space and point out that discovering or excluding them at the LHC can provide important indirect information for a heavy Higgs. Finally, for a very heavy Higgs boson, beyond the search limit at the 7 TeV LHC, we discuss three additional scenarios where models would be consistent with electroweak precision tests: including an additional vector-like fermion mixing with the top quark, adding another $U(1)$ gauge boson and modifying triple-gauge boson couplings. 
\end{abstract}  
  
\thispagestyle{empty}  
\newpage  
  
\setcounter{page}{1}

\baselineskip18pt   

\vspace{-3cm}

\section{Introduction}
\label{sec:intro}
The exciting LHC era will soon answer one of the most important questions in particle physics: the existence or nonexistence of a light Standard Model (SM) Higgs boson. This will  be the most valuable result  in particle physics in the last thirty years. The discovery of a SM Higgs boson will complete the SM and the argument for the existence of new physics will be solely from a naturalness viewpoint. On the other hand,  nonexistence of a SM Higgs boson will be more interesting in a sense that it gives us hints of new particles or new dynamics at the TeV scale. Discovering those additional particles and dynamics in the absence of a SM Higgs boson would be a subsequent focus of the LHC program.

From the viewpoint of simplicity, the Higgs mechanism is an economical way to provide the $W$ and $Z$ gauge boson masses as well as fermion masses in the SM. The Higgs couplings to gauge bosons and fermions are hence dictated by  electroweak symmetry breaking (EWSB) and should not be modified too much from physics at a higher scale. The null result for the SM Higgs from the LHC searches does not immediately lead to the conclusion that no fundamental Higgs field is responsible for EWSB. Actually, there are two generic possibilities to explain the null 7 TeV LHC Higgs searches: the Higgs boson has a new non-standard decay channel that suppresses the branching ratios of the SM decay channel, or the production cross section of the Higgs boson from gluon fusion is suppressed because of other QCD charged particles contributing to the effective operator between the Higgs boson and two gluons.
For sure, another plausible possibility to explain the non-existence of a SM Higgs boson at the LHC would be {\it no Higgs boson} at all and use new strong dynamics like the Technicolor models~\cite{Hill:2002ap} or Higgsless models~\cite{Csaki:2003zu} to explain EWSB. 

Mechanisms to hide the SM Higgs boson is not new at all in the literature. There are numerous activities that concentrate on a light Higgs boson with a mass below 200 GeV (see \cite{Chang:2008cw} for a recent review). However, less attention has been paid to the case of a heavy Higgs boson, which will be the main focus of this paper. One motivation to consider a heavy Higgs boson is that the fine-tuning problem becomes less stringent as for a lighter one~\cite{Barbieri:2006dq}. Another motivation actually comes from the electroweak precision test (EWPT). As is well known, a heavy Higgs boson is not preferred by the electroweak precision data. For example, the oblique parameters $S$, $T$ and $U$~\cite{Peskin:1990zt, Peskin:1991sw}  prefer a lighter Higgs boson, assuming there are no new  contributions. Therefore, a heavy Higgs boson should always be accompanied by new particles beyond the SM to be consistent with the EWPT. It is not hard to imagine that these new particles could change the Higgs properties as well. Taking simplicity as a guidance, in this paper we consider adding only one new particle at a time charged under the SM gauge group  for both to obtain consistency with the electroweak precision observables and to hide a heavy Higgs boson from direct searches.

Considerable efforts have been spent on relaxing the electroweak constraints on the Higgs boson mass, which were summarized into three scenarios in Ref.~\cite{Peskin:2001rw} ten years ago. The first scenario is to add particles whose vacuum polarization integral shifts $S$ in the negative direction. The main example of this is given by scalar fields in several specific multiplets of $SU(2)_W \times SU(2)_c$, where the first $SU(2)_W$ is the weak interaction gauge group and the second one is the custodial symmetry group~\cite{Georgi:1991ci, Dugan:1991ck}. The second method is to add heavy $Z^\prime$ vector bosons to shift all three oblique corrections~\cite{Langacker:1991pg, Holdom:1990xp, Altarelli:1991fk, Altarelli:1993rk, Rizzo:1994ez, Casalbuoni:1998np}. Finally, one could add new particles that produce a nonzero, positive $T$ with or without changing $S$. This have been implemented in quite a few new physics models, for instance, the `topcolor seesaw' where EWSB arises from a heavy $SU(2)_W$ singlet fermion~\cite{Dobrescu:1997nm}. Here, we will loosely follow~\cite{Peskin:2001rw} and introduce new scalars or fermions which are charged under the electroweak gauge group and modify the $S$ and $T$ parameters at the same time. 

Our main focus, however, is to explore how the new physics required by the EWPT modifies the properties of a heavy Higgs,in particular how the Higgs can be hidden at the 7 TeV LHC. If the new particles are also charged under QCD, the production cross section of the heavy Higgs boson from gluon fusion can be modified and even reduced dramatically compared to the SM rate. One such example we will discuss in detail is a colored scalar with a negative quartic coupling to the Higgs. 
After taking into account  the current collider constraints of these new colored scalar particles, we find that a viable model exists to reduce the gluon-fusion production cross section of the Higgs boson by as much as 90\%. Hence, a heavy Higgs boson consistent with EWPT could still be allowed by the 7 TeV LHC searches. Since these colored particles have large production cross sections at the LHC, performing a specific search for these states at the LHC can indirectly provide constraints on a heavy Higgs boson. 

For new QCD-singlet particles, the production cross section can not be modified dramatically, but new decay channels of a heavy Higgs boson can open up. However, this way of hiding a Higgs can only work for a Higgs boson with an intermediate mass below 400 GeV, above which the Higgs boson SM decay width becomes so large that the partial width of the new decay channels could not dominate in any perturbative model. Below, we will check the current collider constraints on these new QCD-singlet  particles and discuss various viable non-standard decays of a heavy Higgs boson. 

Our paper is organized as following. In Section~\ref{sec:stu}, we first review the current status about electroweak precision measurements with an emphasis on the oblique parameters. Then,  we discuss how to hide a heavy Higgs boson by including a new color-singlet particle in Section~\ref{sec:color-singlet}, where we will first check the electroweak precision constrains on the masses of different isospin states in Section~\ref{sec:scalarewpt} and then study the collider signatures as well as constraints in Section~\ref{sec:scalarcollider}. For QCD-charged particles in Section~\ref{sec:QCD}, we first consider the QCD charged scalar and consider their constraints from the EWPT as well as from colliders in Section~\ref{sec:QCDscalars}. The modifications on the Higgs production cross section from gluon fusion will be discussed in Section~\ref{sec:higgsproduction}. We then consider the fermion case in Section~\ref{sec:QCDfermions} by mixing a new fermion with the top quark. After that, we also consider collider constraints on an additional $U(1)$ gauge boson mixing with the $Z$ boson and hence modifying the electroweak precision observables in Section~\ref{sec:zprime}. For the last case of a non-linearly realized EWSB, we discusssed a scenario to transfer the constraints from oblique parameters to triple-gauge boson couplings in Section~\ref{sec:nonlinear}. Finally, we conclude in Section~\ref{sec:conclusion}.

\section{Oblique parameter analysis}
\label{sec:stu}
The usual wisdom to prefer a lighter Higgs boson is because a light Higgs boson is more consistent with the EWPT. Using the recent results from the Gfitter group~\cite{Baak:2011ze}, the Higgs mass is constrained to be $96^{+31}_{-24}$~GeV by the standard fit and $120^{+12}_{-5}$~GeV by a complete fit including the LEP data, the Tevatron and 2010 LHC null results of direct Higgs searches. The upper mass constraint for a SM Higgs boson is 169~GeV (200~GeV) at 95\% (99\%) C.L. from the standard fit and 143~GeV (149~GeV) from the complete fit. The shortly-coming LHC direct serches with a luminosity of 5-10 fb$^{-1}$ should cover all the mass range of a light Higgs boson. 

In many new physics models, additional particles can easily modify the electroweak precision observables.  So, a more proper attitude towards a heavy Higgs boson around or above 200 GeV is to include additional new heavier particles in the EWPT. In this paper, we are going to take this attitude and consider minimal models by including only one new particle at a time. The common approach to constrain physics beyond the SM with the precision electroweak data is through the formalism of oblique parameters: $S$, $T$ and $U$~\cite{Peskin:1990zt, Peskin:1991sw}. The $S$ $(S+U)$ parameter measures new physics contributions to the derivate differences of gauge current vacuum polarizations at zero momenta. The $T$ parameter indicates the difference between the new physics contributions of neutral and charged vacuum polarization at low energies, {\it i.e.}, it is sensitive to weak isospin violation. Generally as the new physics predicts a negligible contribution to $U$ with a few exceptions such as models with anomalous $W$ interactions~\cite{Altarelli:1990zd}, one could fix $U=0$ and only consider the constraints from the $S$ and $T$ parameters.

 \begin{figure}[h!]
\begin{center}
\hspace*{-0.75cm}
\includegraphics[width=0.48\textwidth]{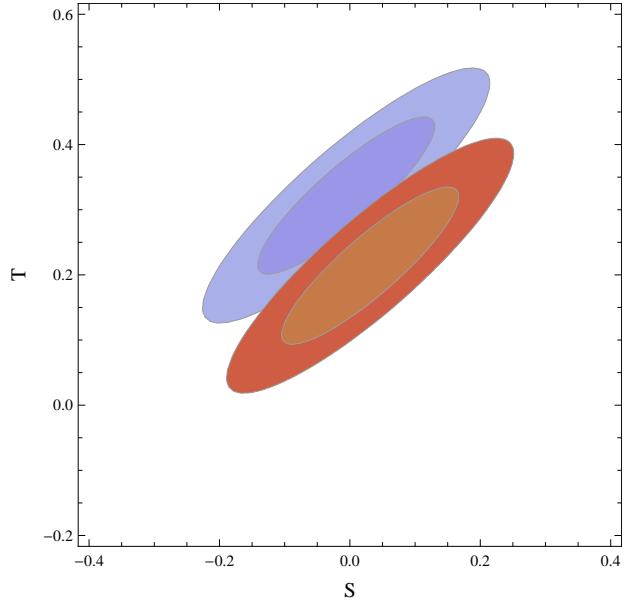} 
\caption{The $S-T$ contour plot with the reference SM Higgs mass at 500 GeV (blue and upper) and 250 GeV (red and lower). For each mass, the two contours correspond to 68\% and 95\% C.L. constraints.}
\label{fig:SMST500250}
\end{center}
\end{figure}

Fixing $U=0$, the most updated global electroweak fit at the reference point $m^{\rm ref}_h = 120$~GeV and $m_t = 173$~GeV is~\cite{Baak:2011ze}
\beqa
S|_{U=0} = 0.07\pm 0.09\,, \qquad\qquad T|_{U=0} = 0.10 \pm 0.08\,, \qquad\qquad (m^{\rm ref}_h = 120~\mbox{GeV}) \,,
\eeqa
with a correlation coefficient of $+0.88$~\footnote{The fit using only low-energy experiment data such as atomic parity violation and lepton scattering prefers a larger value for the $S$ parameter~\cite{Erler:2010sk}. In this paper, we consider the result of fit including also high energy experiment data.}.  Shift in the reference point has to be compensated by shifts in the $S$ and $T$ parameters. For a Higgs boson heavier than 120 GeV, the central value of $S$ from new physics contribution is required to be reduced by $-1/(12\pi) \ln{(m_h^2/120^2)}$ while the central value of $T$ from new physics is increased by $3/(16\pi \cos^2{\theta_W})\ln{(m_h^2/120^2)}$. For instance, at $m^{\rm ref}_h =500$~GeV and $m_t = 173$~GeV, the new physics should have the following contributions to the oblique parameters
\beqa
S|_{U=0} = -0.006\pm 0.09\,, \qquad\qquad T|_{U=0} = 0.32 \pm 0.08\,, \qquad\qquad (m^{\rm ref}_h = 500~\mbox{GeV}) \,.
\eeqa
Thus for theories with a heavy Higgs to be compatible with the precision data, there should be new particles shifting $T$ in the positive direction and/or pushing $S$ to be negative. The allowed regions in the $S$ and $T$ plane with $m^{\rm ref}_h =250$~GeV and $m^{\rm ref}_h =500$~GeV are presented in Fig.~\ref{fig:SMST500250}. One can easily see from Fig.~\ref{fig:SMST500250} that a heavy Higgs boson without other new physics is inconsistent with the electroweak observables at more than 3$\sigma$ level.

In general, one can introduce additional weak charged particles and adjust the mass differences of their isospin components to fix the $T$ parameter while keeping the $S$-parameter almost untouched. Yet the constraints from fitting the $T$ parameter can not set a constraint on the absolute mass scales. Interestingly, by requiring those particles to modify the Higgs decays or production cross sections, one can also fix the masses of those particles and hence have a pretty concrete prediction for the LHC. For sure, this kind of prediction is only possible due to our simplicity assumption that only one new particle is relevant for both EWPT and Higgs phenomenologies.  

\section{Hiding a Heavy Higgs Using a New Color-singlet Particle}
\label{sec:color-singlet}
For color-singlet and $SU(2)_W$ charged particles, the production cross section of the Higgs boson can not be modified significantly. So, in this section we will concentrate on the parameter space where the heavy Higgs boson has a new  decay channel dominant over the SM channels. In principle, the new particles could be scalars or fermions. However, to fix the electroweak precision observables we found that a large mass splitting is required. Thus the fermion case is not preferred as no renormalizable operators can be written down to achieve that and a large modification of the $T$ parameter is not anticipated. On the contrary, renormalizable operators coupling the SM Higgs field to the new weak multiplet exists to generate a sizable splitting inside the scalar multiplet. Therefore, in this section we only study the scalars and consider two models with a weak doublet or a weak triplet. 

\subsection{Electroweak Precision Test}
\label{sec:scalarewpt}

\subsubsection{Scalar Doublet}
\label{sec:scalardoublet}
We first consider the weak doublet model, which has been studied before with an emphasis on the dark matter phenomenology under the name of inert doublet models~\cite{Barbieri:2006dq, Dolle:2009fn}. Here, we will not use the dark matter relic abundance as a constraint on the parameter space but rather consider more generic collider consequences of those new particles. It is shown that these new scalars could produce a negative $S$ as long as the lightest state in the multiplet also has the smallest spin~\cite{Georgi:1991ci, Dugan:1991ck}. In~\cite{Georgi:1991ci, Dugan:1991ck}, the custodial symmetry is always preserved by the interactions of the new electroweak multiplets and hence the $T$ parameter is not modified.  Below we will consider a more general model with custodial breaking operators, in which both $S$ and $T$ will be modified. 

The model contains an addition scalar doublet $\Phi$ transforming as $2_Y$ under $SU(2)_W \times U(1)_Y$. For simplicity, we impose an approximate or exact $\mathbb{Z}_2$ parity on the new doublet and first consider only $\mathbb{Z}_2$ conserving operators. For $Y= 1/2$, this is exactly the inert doublet model considered in \cite{Barbieri:2006dq}. Further studies on this model could be found in~\cite{Dolle:2009fn, Dolle:2009ft, Miao:2010rg}. The scalar potential of Higgs and $\Phi$ is
\beqa
V =  \mu_1 |H|^2 + \mu_2|\Phi|^2+  \lambda_1 |H|^4+ \lambda_2 |\Phi|^4+\lambda_3 |H|^2|\Phi|^2 +\lambda_4 |H^\dagger \Phi|^2+\frac{\lambda_5}{2}\left[ (H^\dagger \Phi)^2+h.c.\right].
\label{eq: doubletpotential}
\eeqa
Notice that the last operator is only present when $Y= 1/2$ and breaks the continuous Peccei-Quinn symmetry $U(1)_\Phi$ enjoyed by the other operators down to the $\mathbb{Z}_2$. The operator with the coefficient $\lambda_4$ splits the masses of components with different isospins while the last operator with the coefficient $\lambda_5$ further breaks the degeneracy between the real and axial neutral scalars. Throughout this paper, we will always assume $\lambda_5$ is small such that the real and axial neutral scalars have approximately equal masses. The potential is bounded from below if and only if
\beqa
\lambda_{1,2}>0; \quad \lambda_3,~ \lambda_3+\lambda_4-|\lambda_5| > -2  \sqrt{\lambda_1\lambda_2}\,.
\eeqa
Under this condition, the minimum with $\langle \Phi \rangle =0$ is stable and the global one provided all the masses of the scalar fields are positive. All the parameters in Eq.~(\ref{eq: doubletpotential}) would be renormalized and the potential stays perturbative up to a reasonably high scale $\sim$ 2 TeV provided the quartic couplings are not too big. Among the quartic couplings, $\lambda_2$ only affects the self-interactions of $\Phi$ and will always taken to be smaller than 1. As we will show, $\lambda_4$ is fixed by the EWPT and is also small $\lesssim 1$.  $\lambda_3$'s beta function is $\beta_{\lambda_3} \sim \lambda_3^2/(4\pi^2)$. We will require $\lambda_3 < 4$ so that the radiative correction to $\lambda_3$ will not exceed 30\% of its tree level value given the cutoff of the model is 2 TeV. 
Parametrizing the $\Phi$ field as $\Phi=(\phi_2,\, \phi_1)^T$ and after EWSB, we have
\beqa
\Delta^2 \equiv m_1^2 - m_2^2 = \lambda_4\,v^2_{\rm EW} \,, \qquad \delta \equiv m_1 - m_2 = \sqrt{m_2^2 +  \lambda_4\,v^2_{\rm EW} } - m_2 \,,
\eeqa
where we neglected the $\lambda_5$'s contribution to those masses in the limit $\lambda_5 \ll \lambda_4$. Here, $m_1$ $(m_2)$ denotes the mass of  $\phi_1$ $(\phi_2)$. $v_{\rm EW}$ is the Higgs vacuum expectation value (VEV) and $ v_{\rm EW} \equiv \langle H \rangle =175$~GeV.

\begin{figure}[!]
\begin{center}
\hspace*{-0.75cm}
\includegraphics[width=0.48\textwidth]{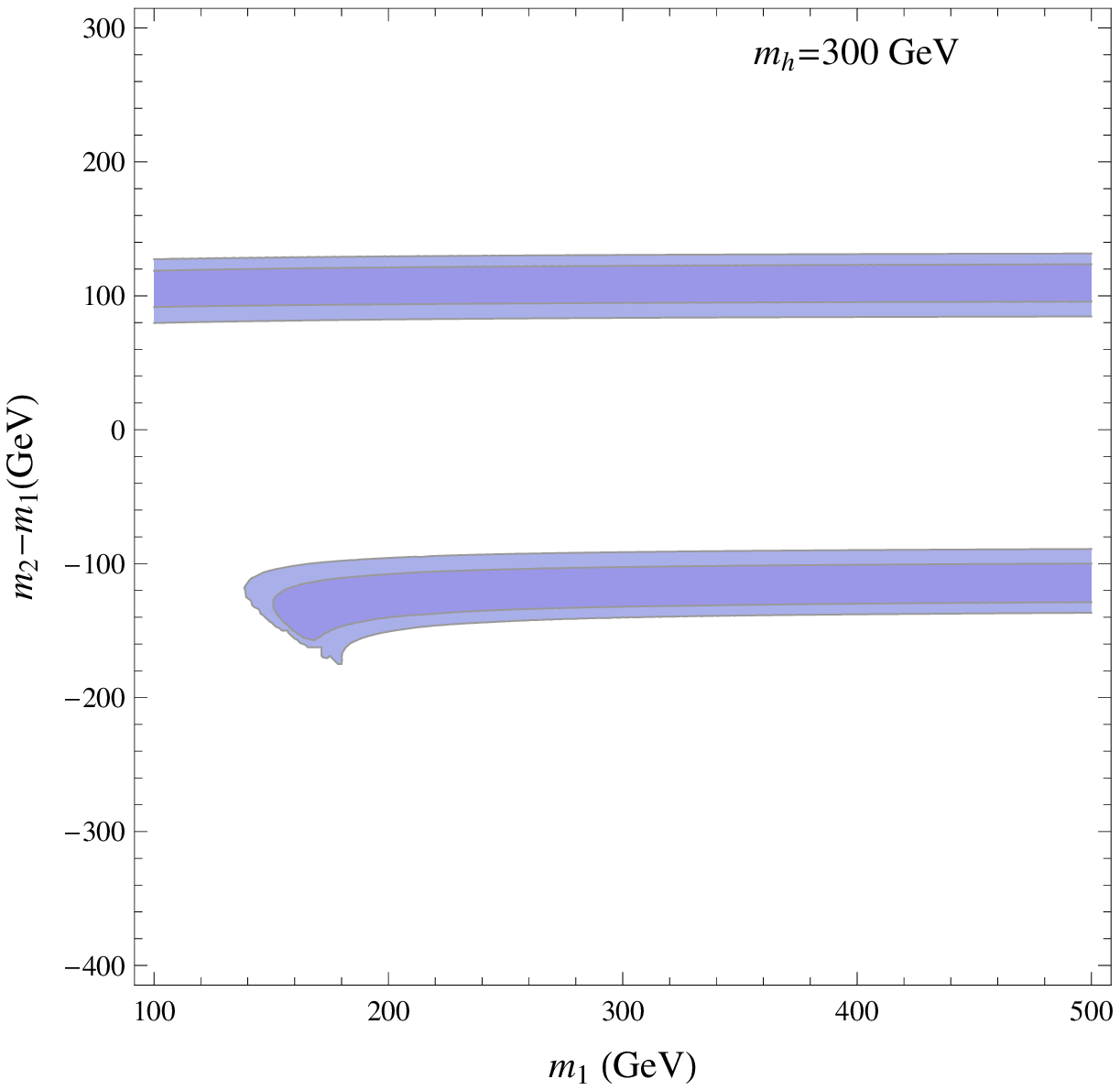}  \hspace{3mm}
\includegraphics[width=0.48\textwidth]{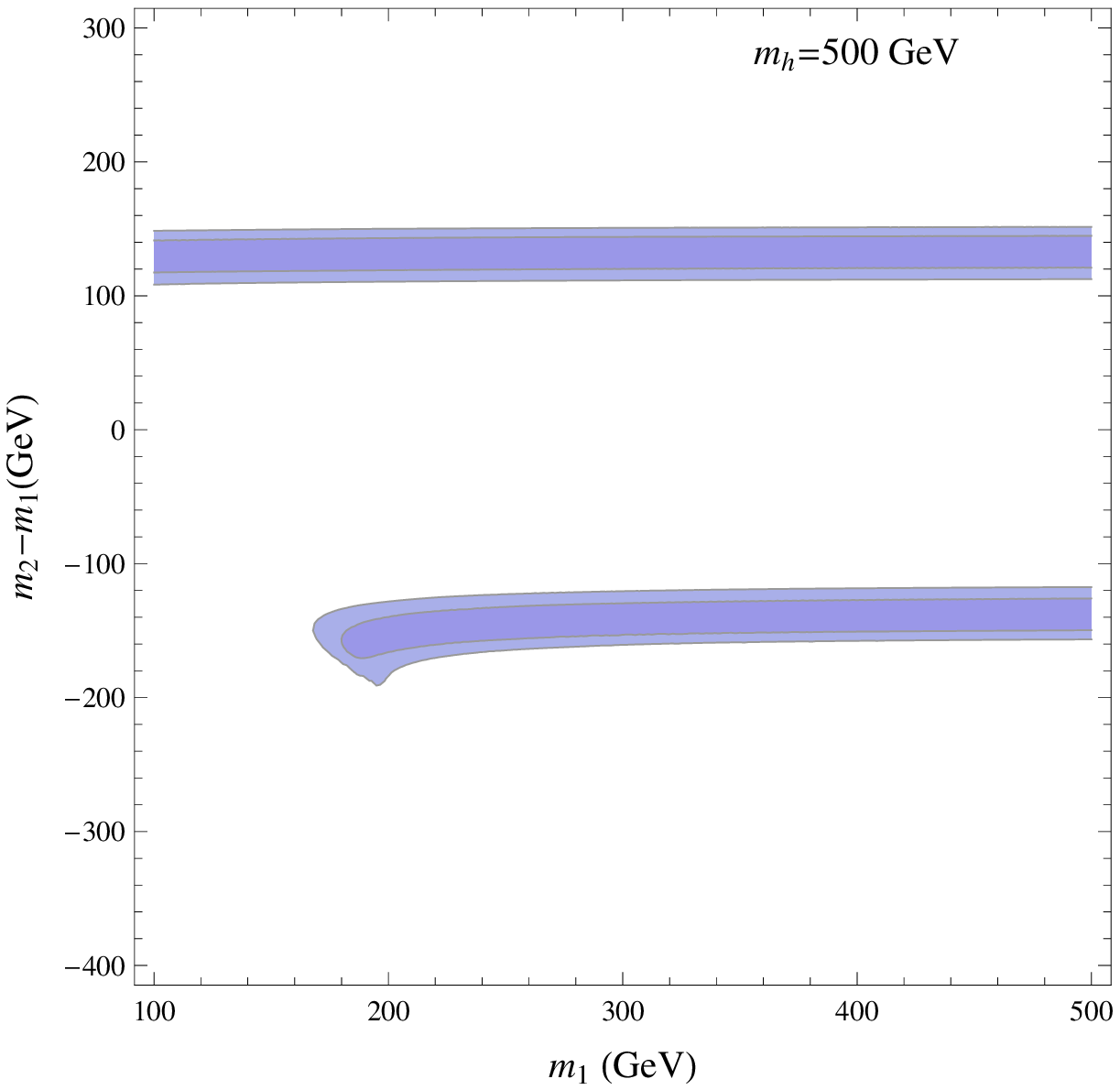}
\caption{The allowed regions in the $(m_1, m_2-m_1)$ plane in the scalar doublet model with $Y=\frac{1}{2}$ from a fit to the $S$ and $T$ parameters. The two contours correspond to 68\% and 95\% C.L. respectively.}
\label{fig:scalarmass}
\end{center}
\end{figure}

In this model, the modification of the $T$ parameter is~\cite{Barbieri:2006dq, Li:1992dt} 
\beqa
\Delta T = \frac{1}{8 \pi s_W^2 M^2_W} F(m_2, m_1) \equiv \frac{1}{8 \pi s_W^2 M^2_W} \left[ \frac{m_1^2 + m_2^2}{2} - \frac{m_1^2 m_2^2}{m_1^2 - m_2^2} \ln \left( \frac{m_1^2}{m_2^2}\right)   \right] \,,
\eeqa
with $F(m_2, m_1) = \frac{2}{3} (m_2 - m_1)^2$ for $m_2 - m_1 \ll m_i$ and $s_W^2 \equiv \sin^2{\theta_{W}} \approx 0.23$. The modification of the  $S$ parameter is~\cite{Barbieri:2006dq, Li:1992dt}~\footnote{The formula for the S parameter in Ref.~\cite{Li:1992dt} is off by a factor of 2 (this is also pointed out in Ref.~\cite{Zhang:2006vt}).}  
\beqa
\Delta S = -\frac{Y}{6 \pi} \ln \left( \frac{m_2^2}{m_1^2} \right) \,.
\eeqa
Choosing $Y=\frac{1}{2}$, we show the constraints on the masses of the two different components of $\Phi$ in Fig.~\ref{fig:scalarmass}. One can see that there are two prefered horizontal bands with the mass splitting around 100 - 140 GeV, which is almost independent of the scalar mass and the heavy Higgs mass. We also checked that the contributions to the $U$ parameter is small for the range of parameter in Fig.~\ref{fig:scalarmass}  and a fit including the $U$ parameter does not modify the conclusion above. 
%

\subsubsection{Scalar Triplet}
\label{sec:scalartriplet}
As  a second example, we consider an electroweak triplet $\Phi$ transforming as a $3_Y$ under $SU(2)_W \times U(1)_Y$~\cite{Hambye:2009pw, Araki:2011hm}. The most generic potential at the renormalizable level is 
\beqa
V =  \mu_1 |H|^2 + \mu_2|\Phi|^2+  \lambda_1 |H|^4+ \lambda_2 |\Phi|^4+\lambda_3|H|^2 |\Phi|^2+\lambda_4(\Phi^\dagger t^a \Phi)^2+\lambda_5(H^\dagger \tau_a H)(\Phi^\dagger t^a \Phi)  ,
\eeqa
where $\tau^a = \sigma^a/2$ with $\sigma^a$ as the Pauli matrices; $t^a$ are the $SU(2)$ generators for spin-1 representation with $t^3={\rm diag}(1, 0, -1)$. For $Y=1$, there is an additional renormalizable operator $\tilde{H}^T \vec{\tau} \cdot \vec{\Phi} H^*+ h.c$, where $\tilde{H}=i\sigma_2 H$ and $\Phi_+ = \Phi^3, \Phi_{++}=\frac{1}{\sqrt{2}}(\Phi^1+i\Phi^2), \Phi_0=\frac{1}{\sqrt{2}}(\Phi^1-i\Phi^2)$. It could be forbidden by a $\mathbb{Z}_2$ symmetry acting on $\Phi$. The physical fields appear in the parameterization of the triplets as follows: $\Phi \equiv (\phi_3, \phi_2, \phi_1)^T$ and each of them has electric charge $Q=T_3+Y$. Only the last operator in the potential splits the masses of different components inside a complex triplet with a non-zero $Y$. For $Y=0$,  this operator vanishes identically as the triplet is real. Thus the real triplet does not contribute to the $S$ and $T$ parameters. From now on, we will only consider a complex triplet with a non-zero $Y$ whose components have masses
\beqa
m_3^2 = m_1^2 + 2 \Delta^2\,, \quad \mbox{and}\,\quad m_2^2 = m_1^2 + \Delta^2 \,,
\eeqa
with $\Delta^2 \equiv \lambda_5\,v_{\rm EW}^2/2$. The condition for the potential to be bounded from below and the existence of a global minimum at $\langle \Phi \rangle = 0$ is 
\beqa
\lambda_{1} >0; \quad \lambda_2 > |\lambda_4|; \quad \lambda_3> -2 \sqrt{\lambda_1\lambda_2}; \quad \lambda_3-\frac{|\lambda_5|}{2}>-2\sqrt{\lambda_1(\lambda_2+\lambda_4)}.
\eeqa
Similar to the doublet model case, we require $\lambda_3\lesssim 4$ to preserve perturbativity up to 2 TeV.  

\begin{figure}[!]
\begin{center}
\hspace*{-0.75cm}
\includegraphics[width=0.48\textwidth]{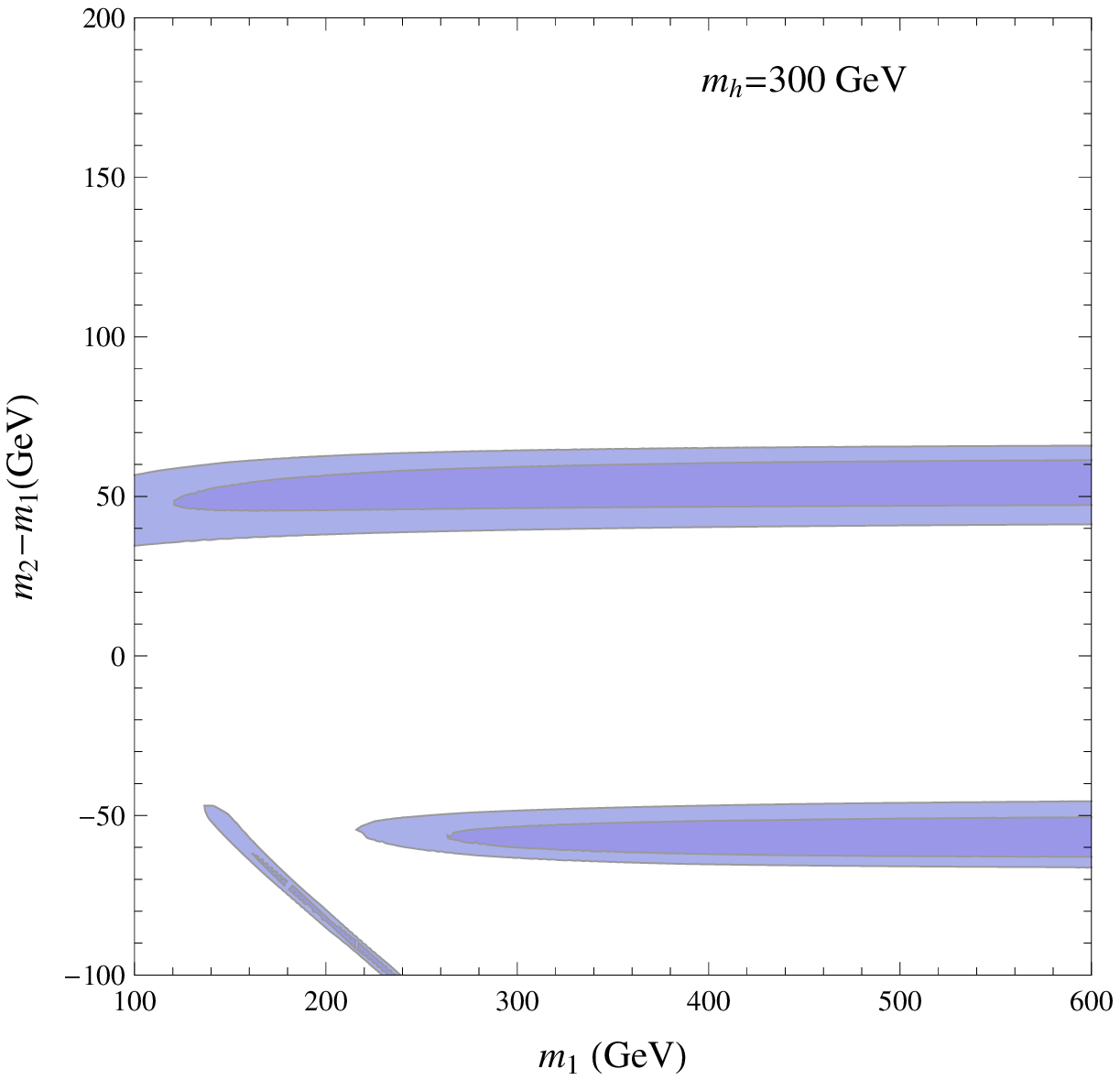}  \hspace{3mm}
\includegraphics[width=0.48\textwidth]{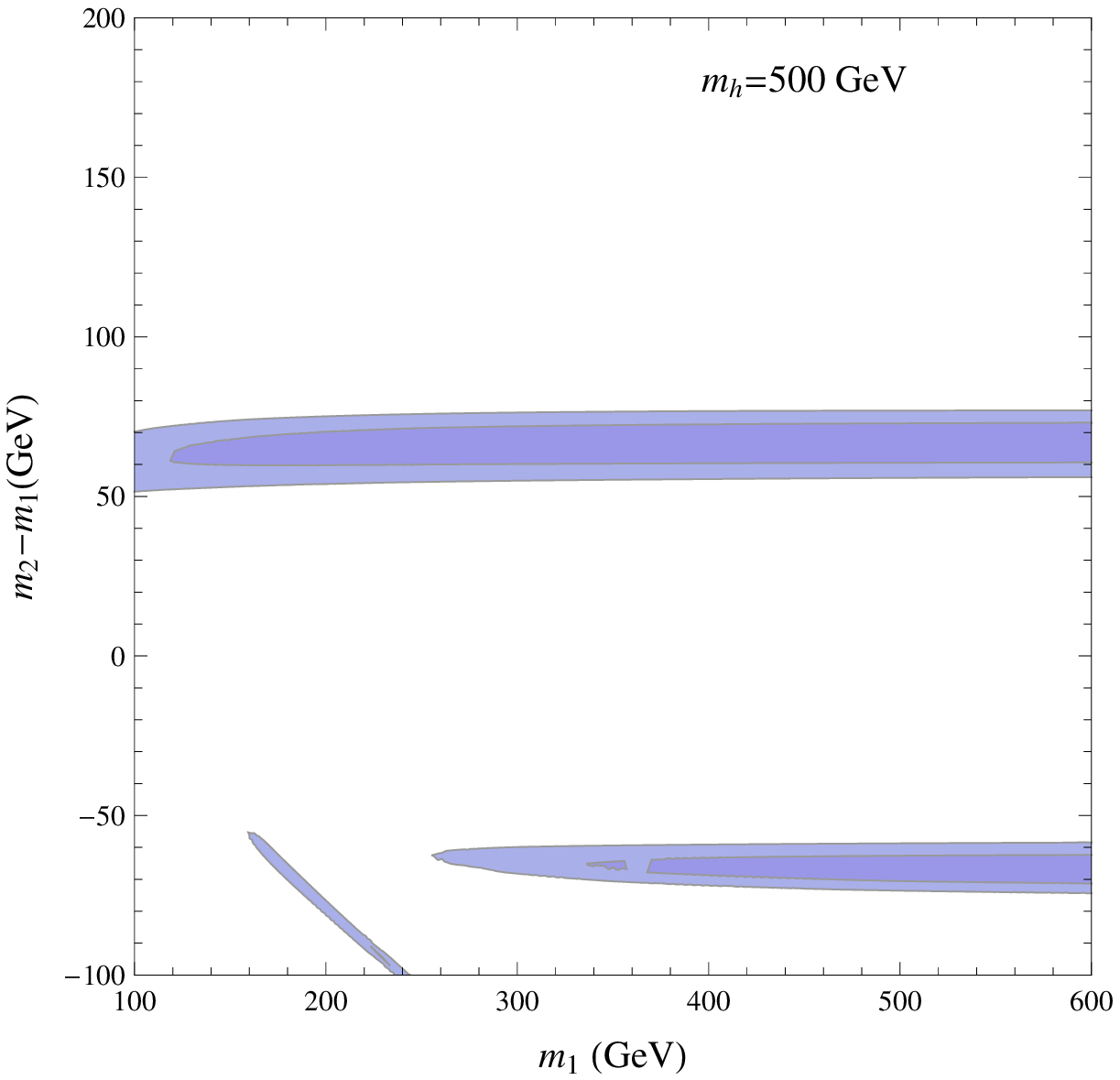}
\caption{The allowed regions in the $(m_1, m_2-m_1)$ plane for a weak-triplet scalar model with $Y=1$ from a fit to the $S$ and $T$ parameters. The two contours correspond to 68\% and 95\% C.L. respectively.}
\label{fig:scalartripletmass}
\end{center}
\end{figure}

The new contribution to the $T$ parameter from the mass splitting of the triplet components is
\beqa
T = \frac{( m_1^2 + \Delta^2 ) \left\{  \Delta^2 \left[  1 - \ln\left( \frac{m_1^2 + 2 \Delta^2}{m_1^2 + \Delta^2} \right)    \right]   - m_1^2\, \tanh^{-1}\left(  \frac{\Delta^2}{m_1^2 + \Delta^2}  \right)  \right\}    }{2\pi s_W^2 M^2_W \Delta^2} \,.
\eeqa
In the small mass splitting limit, we have $T \approx \delta^2/(3\pi\,M_W^2 s_W^2)$ with $\delta \equiv m_2 -m_1$. The contribution to the $S$ parameter from the mass splittings is
\beqa
S = -\frac{Y}{3 \pi} \ln{\left(\frac{m_3^2}{m_1^2} \right) } \,.
\eeqa
In the limit $\delta \ll m_1$, one has $S=-4Y\delta/(3\pi m_1)$. Choosing $Y=1$, we have the allowed regions for $m_1$ and $m_2$ shown in Fig.~\ref{fig:scalartripletmass}. We can see from Fig.~\ref{fig:scalartripletmass} that a triplet with a mass splitting around 50 GeV can be consistent with the EWPT for a heavy Higgs boson.

\subsection{Collider Phenomenologies}
\label{sec:scalarcollider}
The additional scalars charged under $SU(2)_W$ lead to interesting collider signals. They will be produced either indirectly from Higgs decays if kinematically allowed, or they could be paired-produced via weak gauge boson exchanges. The collider signatures highly depend on whether the $\mathbb{Z}_2$ symmetry is broken or not. Below we will first discuss several possibilities of these scalars' decays by coupling them to the SM particles in different ways. Then we will show that they could modify the heavy Higgs decays significantly and thus impact the Higgs searches. We will point out some interesting signatures from the cascade decays of the Higgs boson. Finally, we consider the direct productions of those new scalars and work out the current collider constraints on different decay channels. In this section, we will focus on two benchmark models where scalars have specific hypercharges: the doublet model $2_{1/2}$ and the triplet model $3_1$. 

\subsection{Decays of Scalars}
\label{sec:scalardecay}
If the lightest state inside the scalar $\Phi$ is stable due to the unbroken $\mathbb{Z}_2$ symmetry as in the inert models, it would contribute to the dark matter (DM) density. Thus we have to take the lightest state neutral to avoid the stringent constraints on charged relics. However, unlike the discussions of the inert models, we will not restrict ourselves to the parameter region with the right DM relic abundance. Instead, we will focus on a larger parameter space where the Higgs decay is modified. If the $\mathbb{Z}_2$ is broken by couplings of a single $\Phi$ field to SM fermions and/or gauge bosons, we could have in principle the lightest state to be either electrically charged or neutral. Without loss of generality, we will assume the lightest state to be the electrically neutral one inside the multiplet. 

In the doublet model with $Y=1/2$, $\Phi$ consists of one charged and two neutral particles and can be parametrized as $\Phi= (\phi_+, \phi_0=\left(\phi_r+i \phi_a)/\sqrt{2}\right)^T$. As shown in the previous section, $m_{\phi_+} \approx m_{\phi_0} + $100 GeV from the EWPT. Therefore, the charged state decays to the neutral ones plus an on-shell $W$ gauge boson: $\phi_+ \to \phi_0 + W^+$.  There could be three possibilities of $\phi_0$ decays :
\begin{itemize}
\item{$\phi_r$ or $\phi_a$ is stable. A splitting between $\phi_r$ and $\phi_a$ must be generated by a non-zero $\lambda_5$. Otherwise, $\phi_r$ and $\phi_a$ have an unsuppressed vector-like interaction with the $Z$ boson, which lead to a large spin-independent elastic cross section scattering off nucleus, many orders of magnitude above the current direct detection limit~\cite{Aprile:2011hi}. Notice that this is true even in mass regions where the relic density of $\phi_r$ and $\phi_a$ is small. On the other hand, a non-zero splitting above 1 MeV, the kinetic energy of DM in our galactic halo, is not sufficient to fulfill the inelastic scattering. At colliders, this means the heavier neutral scalar, e.g., the axial one $\phi_a$, would decay to the lighter one plus an off-shell $Z$, $\phi_a \to \phi_r Z^*$. For a small splitting ($\lesssim$ 10 GeV), the decay products from $Z^*$ are soft and could not be triggered on unless a hard jet from initial state radiation is present to boost the decay products~\cite{Bai:2011jg}. The decay length is estimated to be
\beqa
c\tau \sim \frac{240 \pi^3 M_Z^4}{g^4 (m_{\phi_a}-m_{\phi_r})^5} \approx 1.6 \,{\rm m}\left( \frac{200 \, {\rm MeV}}{m_{\phi_a}-m_{\phi_r}}\right)^5,
\eeqa
where $g$ is the $SU(2)_W$ gauge coupling. If the mass splitting is smaller than a few hundred MeV, which means $\lambda_5 < 10^{-3}$,  both neutral scalars are collider stable. 
}
\item{$\phi_0 \to b\bar{b}$. At the renormalizable level, one could write down
\beqa
\lambda_d\,\overline{Q}_L\Phi d_R + \lambda_u\,\overline{Q}_L \tilde{\Phi} u_R + \lambda_\ell\,\overline{L}_L \Phi e_R\,,
\eeqa
where the flavor indexes are not shown. Those operators induce $\phi_0$ to decay into two jets or two leptons depending on the strengthes of Yukawa couplings. To avoid any potential flavor problem, we assume the Yukawa couplings follow the pattern of Minimal Flavor Violation (MFV) to match the SM Higgs Yukawa coupling pattern to fermions.  For $m_{\phi_0} < 2 m_t$, $\phi_0 \to b\bar{b}$ is the dominate decay channel. Notice that these operators break $\mathbb{Z}_2$ parity and induce mass mixing terms such as $\mu^2\,\Phi^\dagger H + h.c.$ in the scalar potential at the one-loop level. Without considering any accidental cancellation between the tree level and the loop-level contributions, the magnitude of $\mu^2$ is estimated from naive dimensional analysis as 
\beqa
\mu^2 \sim \frac{\lambda_t\Lambda^2}{16\pi^2} \sim\lambda_t  (160\, {\rm GeV})^2 \left(\frac{\Lambda}{2\,{\rm TeV}}\right)^2.
\eeqa 
This radiative contribution would mix $H$ and $\Phi$ and modify the spectrum. To avoid a large mixing between $H$ and $\Phi$, we require the Yukawa couplings to be small, $\lambda_t < 10^{-2}$. Thus the heavier state $\phi_+$ decaying to two SM fermions are suppressed and has a smaller width compared to the decay into the neutral states plus the $W$ gauge boson. If $\lambda_b \lesssim 10^{-8}$, the decay length is of order meters and the lightest state is collider stable.  }
\item{$\phi_0$ decays to two gauge bosons through dimension-six operators 
 \beqa
\frac{c_g( \Phi^\dagger H) G_A^{\mu\nu}G^A_{\mu\nu}}{\Lambda^2} +  \frac{c_w(\Phi^\dagger H) W_a^{\mu\nu}W^a_{\mu\nu} }{\Lambda^2}+ \frac{c_b( \Phi^\dagger H) B^{\mu \nu} B_{\mu \nu}}{\Lambda^2} +\frac{c_{wb}( \Phi^\dagger \sigma^a H) W^a_{\mu\nu} B^{\mu\nu}}{\Lambda^2},
 \eeqa
where $G, W, B$ are field strengths of SM gauge groups. More operators can be written down with covariant derivatives, which may lead to the similar final sates. From those operators, one could have
\beqa
\phi_0 \to gg, \gamma \gamma, Z \gamma. 
\eeqa 
However, this is not the whole story. At the one loop order, all these operators would generate $\mu^2\,\Phi^\dagger H+h.c.$, which by naive dimension analysis is of order $\mu^2 \sim c\,(160 \, {\rm GeV})^2 \left(\frac{\Lambda}{2\, {\rm TeV}}\right)^2$ with the parameter $c$ including various powers of SM gauge couplings as well as the coupling of $\Phi$ to new particles which generate these dimension six operators. To avoid the case that $\Phi$ develops a very large VEV, we assume a very tiny $c$ here. The induced mixing between $\Phi$ and $H$ would then cause light $\phi_0$ decaying to $b\bar{b}$ pair with a partial width estimated to be
\beqa
\Gamma(\phi_0 \to b\bar{b}) = \frac{3y_b^2\mu^2 m_{\phi_0}}{8\pi m_h^4} \sim \frac{3c^2\Lambda^4m_{\phi_0}}{2048\,\pi^5 m_h^4}\,,
\eeqa
with $y_b$ as the SM Higgs coupling to the bottom quark. The ratio between $\Gamma(\phi_0 \to b\bar{b})$ and the width of $\phi_0$ decaying to two gauge bosons, e.g., $\Gamma(\phi_0\to gg)$, scales as 
\beqa
\frac{\Gamma(\phi_0 \to b\bar{b})}{\Gamma(\phi_0 \to gg)} \sim \frac{ y_b^2\Lambda^8}{(16\pi^2)^2 m_h^4 m_{\phi_0}^2 v^2} \approx 3 \left( \frac{\Lambda}{2\,{\rm TeV}}\right)^8 \left( \frac{100\,{\rm GeV}}{m_{\phi_0}}\right)^2\left( \frac{300\,{\rm GeV}}{m_h}\right)^4,
\eeqa
where the dependences on the coefficient $c$ cancel out. Yet, one should bear in mind that there could be large uncertainties in this evaluation by ignoring the inputs of UV physics. If the effective cutoff is lowered $\sim$ 1 TeV, the estimate above leads to comparable branching ratios. Thus, we still keep $\phi_0$ decaying to two gauge bosons as one possibility.}
\end{itemize}

For the triplet case with $Y=1$, $\Phi$ consists of a doubly charged state, a single charged state and a complex neutral state $\Phi\equiv(\phi_{++}, \phi_+, \phi_0)$. The single charged state has a mass $m_{\phi_+} \approx m_{\phi_0} +$ 50 GeV from Fig.~\ref{fig:scalartripletmass} and decays as $\phi_\pm \to \phi_0 W^*$. The doubly charged state is even heavier, $m_{\phi_{++}} = \sqrt{2m_{\phi_+}^2-m_{\phi_0}^2}$, which gives a mass difference $m_{\phi_{++}} - m_{\phi_+}$ smaller than 50 GeV. Thus the doubly charged state also decays to an off-shell $W$ with the single charged state, $\phi_{\pm\pm} \to W^{\pm(*)}\phi_{\pm}$. Analogous to the neutral state in the doublet model, there are three possibilities for $\phi_0$ decays:
\begin{itemize}
\item{one component of $\phi_0$ is stable. To avoid the constraints from DM direct detection experiments, we need to include a dimension six operator $(\tilde{H}^T \vec{\tau} \cdot \vec{\Phi} H^*)^2 + h.c.$ to split the real and axial components of the neutral scalar. This mass splitting could be naturally small $\delta \sim v_{\rm EW}^4/(2\Lambda^2m_{\phi_0})$, which is about 1 GeV for $m_{\phi_0} =$ 100 GeV and $\Lambda \sim$ 2 TeV. Again the axial component can decay into the real one, which could be a stable particle, plus an off-shell $Z$ gauge boson.}
\item{$\phi_0 \to b\bar{b}$ mediated by an operator $\lambda\, \bar{Q}_L  \vec{\tau} \cdot \vec{\Phi}  H^\dagger d_R$, assuming MFV and $m_{\phi_0} < 2m_t$. There are two other similar operators $\bar{Q}_L  H  \vec{\tau} \cdot \vec{\Phi} ^\dagger u_R, \bar{L}_L  \vec{\tau} \cdot \vec{\Phi} H^\dagger e_R$. To avoid large radiative generated term, $\tilde{H}^T \vec{\tau} \cdot \vec{\Phi} H^*+ h.c.$, we require the Yukawa couplings to be small. 
}
\item{$\phi_0$ decays to two gauge bosons through dimension six operators such as
\beqa
(\tilde{H}^T \vec{\tau} \cdot \vec{\Phi} H^*) G_A^{\mu\nu}G^A_{\mu\nu}\,, \quad \,(\tilde{H}^T \vec{\tau} \cdot \vec{\Phi} H^*) W_a^{\mu\nu}W^a_{\mu\nu}\,, \quad  \,(\tilde{H}^T \vec{\tau} \cdot \vec{\Phi} H^*)B^{\mu\nu}B_{\mu\nu}\,.
\eeqa}
\end{itemize}

\subsubsection{Higgs decays}
\label{sec:higgsdecay}
The existence of additional weak-scale scalars opens up new Higgs decay channels. The partial decay widths of Higgs to additional scalars depend on the quartic couplings between two Higgs and two $\Phi$'s in the potential. 

First we consider the doublet model with $Y=1/2$. The partial width of Higgs decaying to the scalars is given by
\beqa
\Gamma(h \to \phi_i \phi_i)= \frac{v_{\rm EW}^2}{16\pi m_h} \lambda_i^2\left(1-\frac{4m_i^2}{m_h^2}\right)^{1/2}\,,
\eeqa
where $i=+, r, a$ and 
\beqa
\lambda_+=\sqrt{2}\lambda_3 \quad \lambda_r=\lambda_3+\lambda_4+\lambda_5 \quad \lambda_a=\lambda_3+\lambda_4-\lambda_5.
\eeqa
For the triplet case, we have
\beqa
\Gamma(h \to \phi_i^\dagger \phi_i)= \frac{v_{\rm EW}^2}{8\pi m_h}c_i^2\left(1-\frac{4m_i^2}{m_h^2}\right)^{1/2},
\eeqa
where $i=1,2,3$ and $c_{1,2,3}=(\lambda_3-\lambda_5/2, \lambda_3, \lambda_3+\lambda_5/2)$. 

\begin{figure}[!h]
\begin{center}
\hspace*{-0.75cm}
\includegraphics[width=0.48\textwidth]{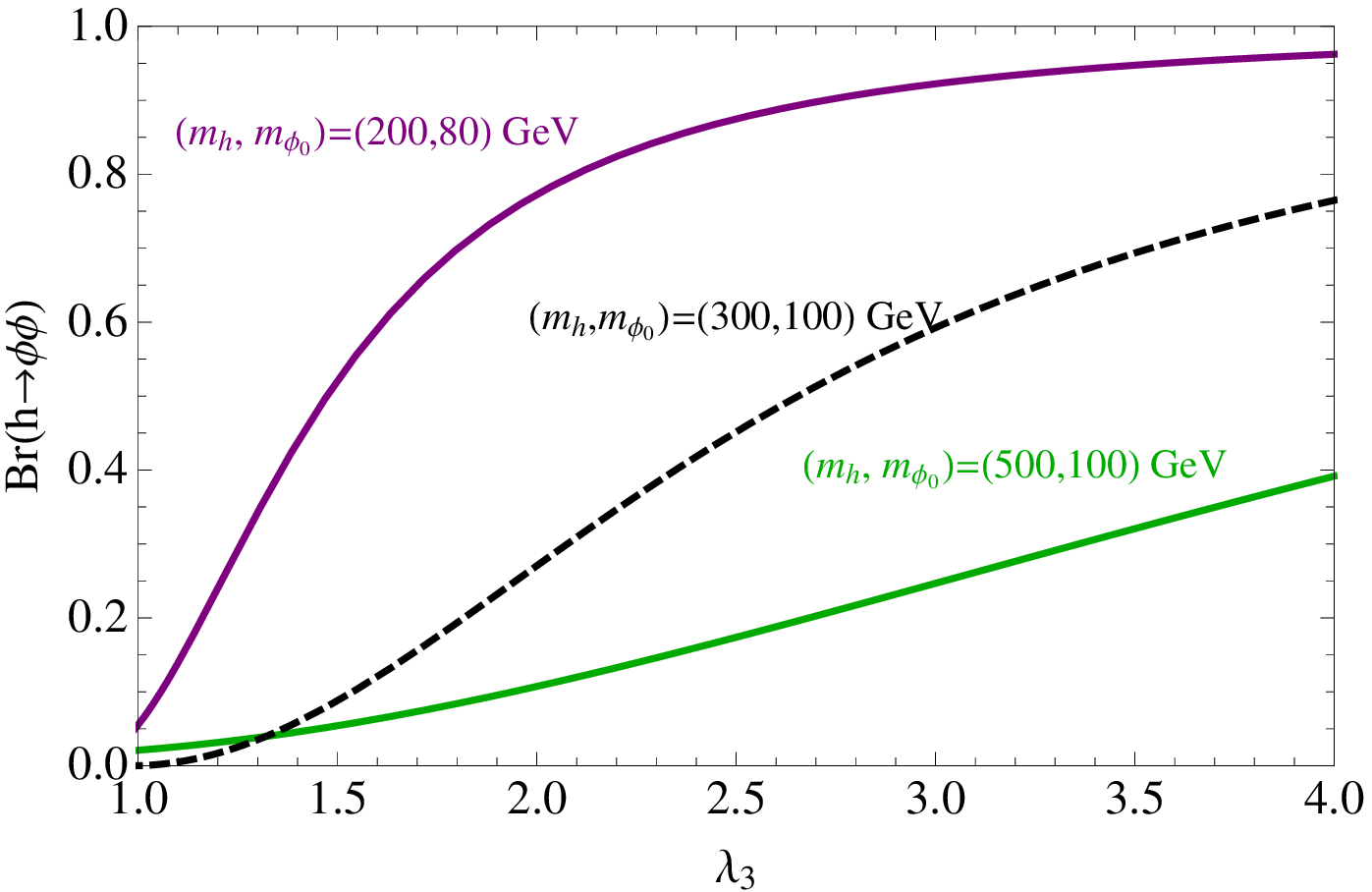} \hspace{3mm}
\includegraphics[width=0.48\textwidth]{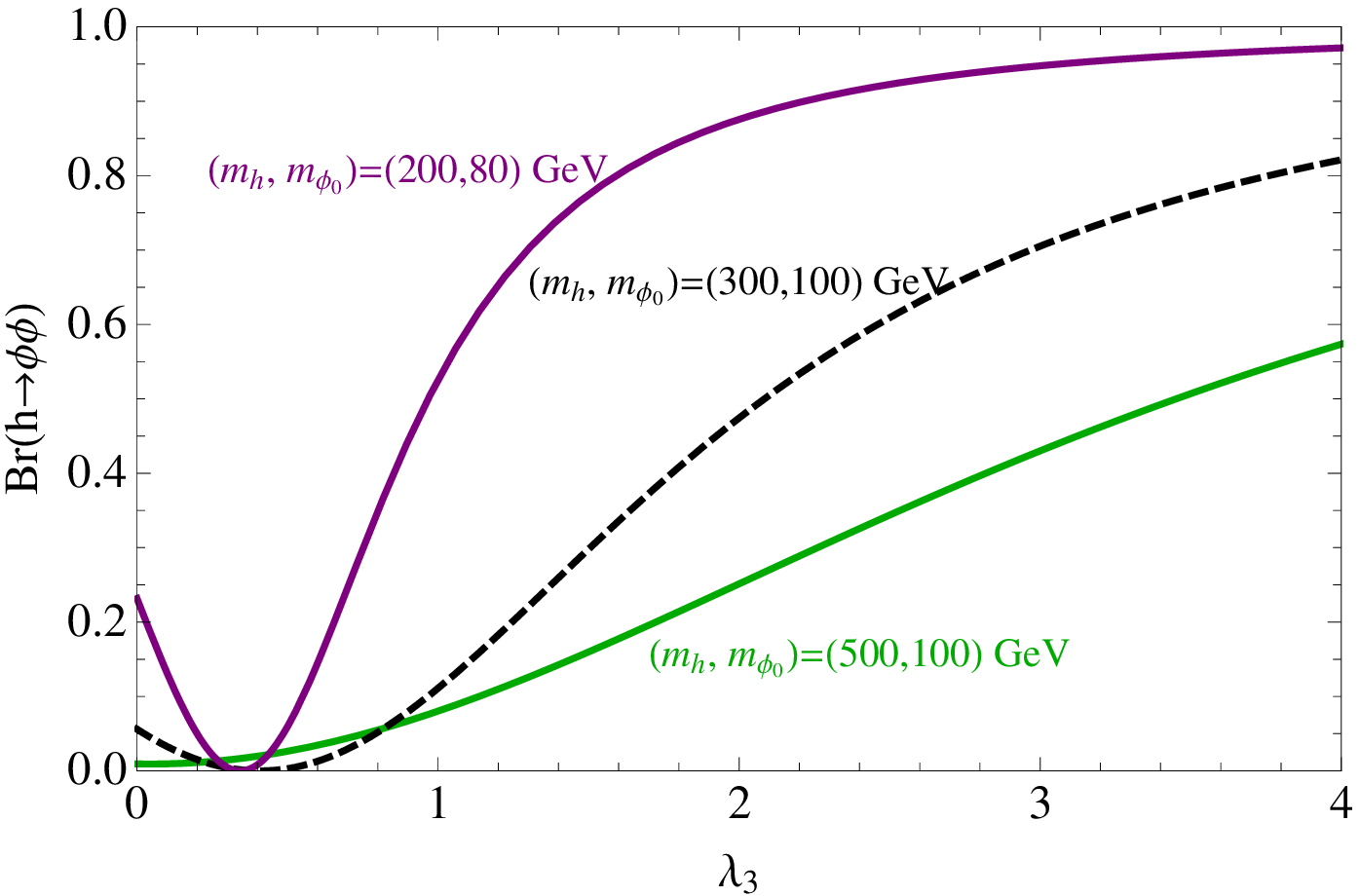}
\caption{Decay branching ratio of the Higgs to the lightest component of an additional doublet (left) and triplet (right) as a function of the coefficient $\lambda_3$ which doesn't contribute to the mass splitting of different components of $\Phi$. For each curve, we fix the Higgs mass $m_h$ and the lowest component mass $m_{\phi_0}$. The mass splitting is $\delta$ =100 GeV (left); $\delta$ =50 GeV (right). }
\label{fig:br}
\end{center}
\end{figure}

The branching fractions of $h \to \Phi \Phi$ are presented in Fig.~\ref{fig:br}. The coefficients that give rise to the mass splitting are fixed by EWPT and we plot the branching faction as a function of the remaining coefficient ($\lambda_3$ in both cases) that preserves the custodial symmetry. From Fig.~\ref{fig:br}, one can see that for Higgs in the mass range 200 - 300 GeV, the Higgs decaying to the new scalars could easily dominate over the Higgs decaying to $2 W$'s or $2 Z$'s, e.g., ${\rm Br}(h \to W^+ W^-) + {\rm Br}(h \to 2Z) \lesssim 0.5$. For an even heavier Higgs boson, the width/mass ratio of Higgs becomes order of unit if one add new decay channels to suppress the SM branching ratios. Therefore, we only concentrate on the intermediate mass ranges in this section.  In the mass range $m_h \in (200, 300)$ GeV, the current Higgs searches with 1 - 2.3 ${\rm fb}^{-1}$ data exclude $\sigma(h\to WW/ZZ) \gtrsim 0.5 \times \sigma_{SM}$~\cite{higgscombination}. It is projected that 5 ${\rm fb}^{-1}$ data could push the limit down to $\sigma(h\to WW/ZZ) \sim 0.4 \times \sigma_{SM}$. As we can see from Fig.~\ref{fig:br}, a lot of parameter space associated with the new scalar particles exist to reduce $\sigma(h\to WW/ZZ)$ and hide a heavy Higgs boson at the 7 TeV LHC.

Although the Higgs boson can be hided in the existing searches by adding a new weak-charged scalar, new signatures from Higgs decays are predicted at the 7 TeV LHC. Taking into account of different $\phi_0$ decays, we could have several interesting possibilities
\beqa
&&h \to \phi_0 \phi_0 \nonumber \\
&&h \to \phi_0 \phi_0 \to 4b, 4g \nonumber \\
&&h \to \phi_0 \phi_0 \to 4 \gamma, 2Z + 2\gamma \,.\nonumber
\eeqa
The first one is the Higgs invisible decay, which could be searched for in the monojet channel, the $Z$ plus missing energy channel and two forward jets plug missing energy channel from $W$-boson-fusion productions. The second and the third possibilities, to hide Higgs in four jets or ``bare'' them in four photons have already been discussed in the context of hiding light Higgs~\cite{Dobrescu:2000jt, Chang:2006bw, Chang:2008cw}. Notice that in the context of hiding light Higgs, the intermediate particles are always very light pesudo-scalars and the final jets or photons could be boosted and collimated while in our scenario, as $\phi_0$ is not very light, the final state particles are not necessarily close to each other. In the triplet model, there could be a small region of parameter space where $h \to \phi_+ \phi_- \to 2 W^{*} 2\phi_0$.

\subsubsection{Direct collider searches}
\label{sec:constraints}
The neutral states $\phi_0$ could not be lighter than around 45 GeV; otherwise, the $Z$ boson could decay to them and the total $Z$ boson width will be modified, which is highly constrained. For a lighter $\phi_0$ with a mass below 100 GeV but above 45 GeV, they could be paired-produced directly at LEP with a cross section~\cite{Barbieri:2006dq}
\beqa
\sigma(e^+e^- \to \phi_r \phi_a)=\left(\frac{g}{2\cos\theta_w}\right)^4\left(\frac{1}{2}-2\sin\theta_w^2+4\sin\theta_w^4\right)\frac{1}{48\pi s}\frac{(1-4m_{\phi_0}^2/s)^{3/2}}{(1-m_Z^2/s)^2} \,,
\eeqa
where $g$ is the weak coupling constant and $\theta_w$ is the weak angle. For center of mass energy $\sqrt{s} = 200$ GeV, $m_{\phi_0}=$ 60 GeV, $\sigma = 0.25$ pb. If $\phi_0$ is stable, it would lead to the mono-photon + MET signal $e^+e^- \to \phi_r \phi_a \gamma$. However, the cross section is small with $p_T(\gamma) > 0.0375 \sqrt{s}$, e.g., $\sigma \approx 0.01$ pb for $m_{\phi_0}=$ 50 GeV, which is beyond the sensitivity of LEP experiments~\cite{LEPmonophoton}. However, if $\phi_0$ decays to SM particles, LEP results put stringent constraints on the parameter space that is kinematically accessible. If $\phi_0$ decays to $b\bar{b}$ 100\% of times, the 4$b$ jet final state search with both LEP1 and LEP2 data rules out $m_{\phi_0}$ up to 90 GeV~\cite{LEP4b}. More concretely, the 4$b$ jet search conducted by the DELPHI collaboration rules out a rate as large as that of paired-production of CP odd state $A$ and Higgs in a two Higgs doublet model in the Higgs mass range from 40 to 90 GeV (see Fig. 11 in~\cite{LEP4b}), assuming $m_A = m_h$; $\cos^2(\alpha-\beta)=1$ and 100\% branching into 4$b$. In our case, the cross section of the 4$b$ final state is the same as $\sigma (e^+e^- \to Ah)$ in the doublet model and three times larger in the triplet model. Similarly, if $\phi_0$ decays to two photons, $\phi_0 \to 2\gamma$, $m_{\phi_0}$ below 90 GeV is ruled out by multi-photon searches at LEP~\cite{LEPmultiphoton}.

\begin{figure}[!h]
\begin{center}
\hspace*{-0.75cm}
\includegraphics[width=0.48\textwidth]{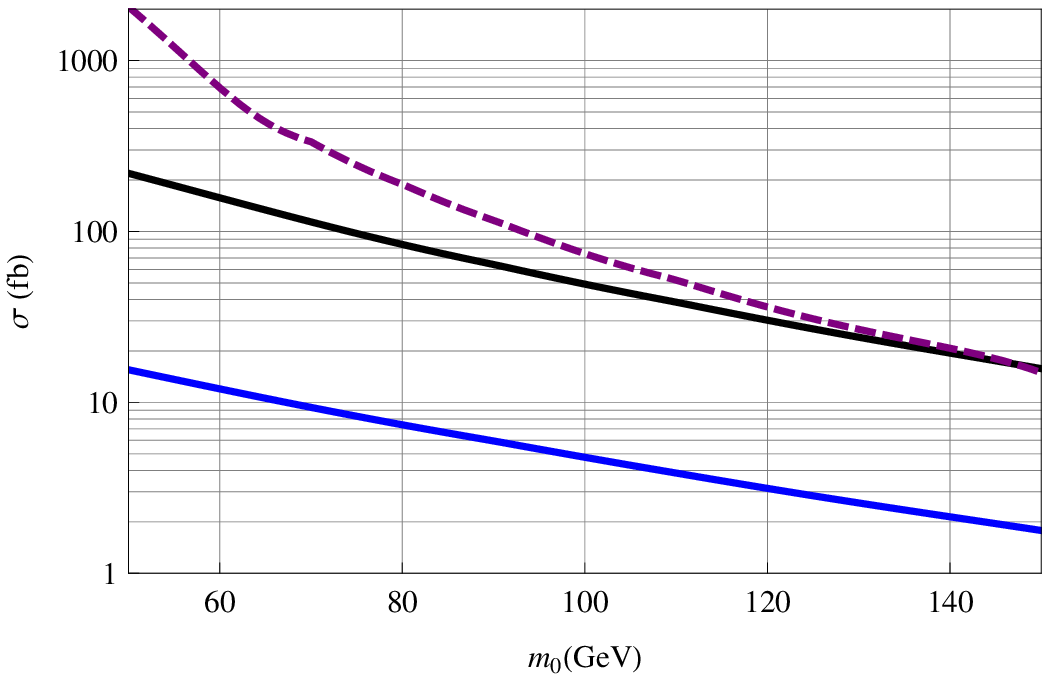}  \hspace{3mm}
\includegraphics[width=0.48\textwidth]{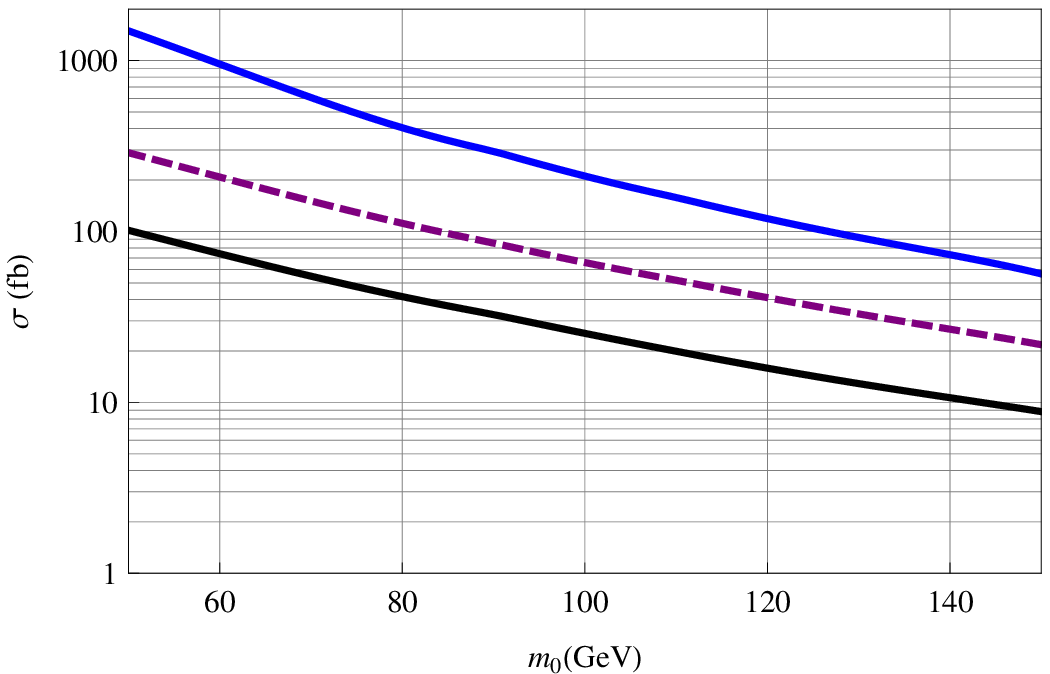}
\caption{Production cross sections of an additional doublet/triplet $\Phi$ at the LHC with $\sqrt{s}=7$ TeV where $m_0$ is the mass of the lightest component of $\Phi$. For a doublet, $\delta m =$100 GeV; for a triplet, $\delta m=$50 GeV. Left: $\phi_0\phi_0$ (purple and lower), $\phi_\pm \phi_0$ (black and middle) and $\phi_+\phi_-$ (blue and upper). Right: $\phi_{\pm\pm}\phi_{\mp\mp}$ (black and lower), $\phi_{\pm\pm}\phi_{\mp}$ (purple and middle) and $\phi_\pm \phi_0$ (blue and upper).}
\label{fig:cs}
\end{center}
\end{figure}

At the hadron colliders, all states of $\Phi$ could be produced through electroweak interactions. For the scalar doublet,
\beqa
pp(\bar{p}) &\to& W^{\pm*} \to \phi_\pm \phi_0 \,, \\
pp(\bar{p}) &\to& Z^{*} /\gamma^{*} \to \phi_+ \phi_- , \phi_r\phi_a   \,,
\eeqa
while for the scalar triplet,
\beqa
pp(\bar{p}) &\to& W^{\pm*} \to \phi_{\pm\pm}\phi_{\mp},  \phi_\pm \phi_0  \,, \\
pp(\bar{p}) &\to& Z^{*} /\gamma^{*} \to  \phi_{++}\phi_{--},  \phi_+ \phi_- , \phi_r\phi_a   \,.
\eeqa
The production cross sections for different channels at the LHC with $\sqrt{s}$ =7 TeV are presented in Fig.~\ref{fig:cs}. The final states and possible signals are categorized in Table.~\ref{table:doubletfs} and~\ref{table:tripletfs}.

\begin{table}[h]
\renewcommand{\arraystretch}{1.3}
\begin{center}
\begin{tabular}{|c|c|c|}
\hline 
& Relevant final states & Possible signals\\
\hline
$\phi_0$ stable & MET, $W$ + MET, $W^+W^-$+MET  & mono-jet+MET, jets+MET, 1 $l$+MET\\
$\phi_0 \to 2 b/g$'s& 4 $j$,  $W + 4 j$, $W^+W^- + 4 j$ & 4 $j$, $W$+$j$'s, OS+MET, 1 $l$ + $b$ jets + MET \\
$\phi_0 \to 2 \gamma/\gamma+Z$ & $n W$'s + $m Z$'s + $l \gamma$'s & SS, multi-leptons, multi-photons, multi-jets \\
\hline
\end{tabular}
\caption{ A sample of collider signals from producing doublet $\Phi$ at hadron collider.}
\label{table:doubletfs}
\end{center}
\end{table}

\begin{table}[h]
\renewcommand{\arraystretch}{1.2}
\begin{center}
\begin{tabular}{|c|c|c|}
\hline 
& Relevant final states&Possible signals\\
\hline
$\phi_0$ stable & MET, $n W$'s + MET, & mono-jet+MET, OS+MET, SS+MET, multi-leptons, multi-jets \\
$\phi_0 \to 2 b/g$'s& 4 $j$,  $W$'s+ 4$ j$, &OS+MET, SS+MET, multi-leptons, multi-jets\\
$\phi_0 \to 2 \gamma/\gamma+Z$ & $n W$'s + $m Z$'s + $l \gamma$'s  &multi-photons, OS+MET, SS+MET, multi-leptons, multi-jets \\
\hline
\end{tabular}
\caption{Collider signals from producing triplet $\Phi$ at hadron collider.}
\label{table:tripletfs}
\end{center}
\end{table}

Given the small electroweak production rates, we found that most of the current searches are not sensitive to these new scalars. For instance, one would worry about constraints on the production of $\phi_{++}\phi_{--}$ and $\phi_{\pm\pm}\phi_{\mp}$ from the same-sign (SS) lepton searches at both Tevatron and the LHC. For instance, the Tevatron SS search adopts a set of very loose cuts~\cite{cdfss}
\beqa
&& \mbox{at least two SS leptons } \nonumber \\
&& \mbox{the leading lepton having } p_T > 20~\mbox{GeV},\; |\eta| < 1.1; \nonumber \\
&& \mbox{the sub-leading lepton having } p_T > 10~\mbox{GeV},\; |\eta| < 1.1; \nonumber \\
&& \mbox{remove the regions } 86~\mbox{GeV} < m_{\ell^+ \ell^-}, m_{\ell^\pm \ell^\pm}  < 96~\mbox{GeV}\; \mbox{and} \;
   m_{\ell^\pm \ell^\pm} < 25~\mbox{GeV}\,,
    \eeqa
which already imposes a strong constraint on the doubly-charged scalar inside an electroweak triplet to have a mass above 245 GeV. However, this limit is set by assuming 100\% branching ratio of $\phi_{++}$ to $ee, \mu\mu$ or $e \mu$. In our case, however, the leptons are from the (off-shell) $W$ decays in the long cascade decay chain $\phi_{++} \to W^{+*} \phi_{+} \to W^{+*}W^{+*} \phi_0$. Thus the cross section of the SS final state is reduced by a factor of $2 \times0.2^2 = 0.08$, where 0.2 is the $W$ leptonically decay branching fraction and the factor 2 takes into account that SS leptons could come from either decay chain in the $\phi_{++}\phi_{--}$ pair production. Besides, the invariant mass of the two SS leptons does not reconstruct  a bump at $m_{\phi_{++}}$.  Thus we conclude that the region $m_{\phi_{++}} > 130$ GeV, or equivalently, $m_{\phi_0} >50$ GeV is not constrained. The CMS SS searches require a much stronger set of cuts~\cite{CMSdilepton}. The preselection cuts are
\beqa
&& p_T(\mbox{jet}) > 40~\mbox{GeV}\; \mbox{with}\;|\eta|< 2.5; \mbox{at least two jets};  \nonumber \\
&& \mbox{two same-sign leptons with }p_T(\mbox{muon}) > 5~\mbox{GeV and } p_T(\mbox{electron}) > 10~\mbox{GeV};
\label{eq:CMSSScuts}
\eeqa
Among the final four signal regions, the search region with low $H_T$ but high $\missET$ cuts, $H_T >$ 80 GeV and $\missET> 100$ GeV is most sensitive to the case where $\phi_0$ is stable and contributes to the missing energy. The high $H_T$ low $\missET$ search region is most sensitive to an unstable $\phi_0$ (decaying to two jets). We used the FeynRules package~\cite{Christensen:2008py} to generate our new physics models and then feed them into MadGraph 5~\cite{Alwall:2011uj} which calculated the matrix elements and simulated events. The events are then showered using Pythia 6.4~\cite{Sjostrand:2006za}. For a stable $\phi_0$, we found that the acceptance of the signal is 7\% for $(m_ {\phi_0}, m_{\phi_{++}})=$ (50, 132) GeV and there are 2 SS events after cuts for 1 ${\rm fb}^{-1}$ luminosity. For $\phi_0 \to b\bar{b}$, we found that the acceptance of the signal could be as large as 50\% for $(m_ {\phi_0}, m_{\phi_{++}})=$ (100, 187) GeV, which yields 3 events at 1 ${\rm fb}^{-1}$ luminosity. They are below the observed upper limits on event yields from new physics~\cite{CMSdilepton}, which is 7.5 for $(H_T^{\rm min}, \missET) = (400, 50)$ GeV and 6 for $(H_T^{\rm min}, \missET) = (80, 100)$ GeV.

In summary, if $\phi_0$ is stable at the collider scales, there is no constraints for the mass regions we are interested in $m_{\phi_0} \in (50, 150)$ GeV. If $\phi_0$ decays promptly to the SM final states, the allowed mass region shrinks to $m_{\phi_0} \in (100, 150)$ GeV due to the LEP constraints. Although the current LHC searches with 1 ${\rm fb}^{-1}$ have not provided a constraint on the models we are considering here, more data would allow us to probe the parameter space in interest and to close this way of hiding Higgs. Especially, the SS lepton searches could set interesting limits on the triplet model very soon.

\section{Hiding a Heavy Higgs using a New QCD-Charged Particle}
\label{sec:QCD}
\subsection{QCD Charged Scalars}
\label{sec:QCDscalars}
Restricting ourselves to the fundamental and adjoint representations of $SU(3)_c$ and $SU(2)_W$, we have four choices of scalars: $(8, 2)_Y$, $(8, 3)_Y$, $(3, 2)_Y$ and $(3, 3)_Y$. Depending on the hypercharges of those scalars, we have different consequences for the electroweak precision observables and couplings to SM particles. For the color-octet scalars, we consider $\mathbb{O}_2\equiv(8, 2)_{1/2}$ and $\mathbb{O}_3\equiv(8, 3)_{1}$ as two examples. The former was considered in Ref.~\cite{Manohar:2006ga} as the only choice other than $(2, 1)_{1/2}$ to realize the MFV in the quark sector at the renormalizable level. For the color-triplets, we choose $\mathbb{T}_2\equiv(3, 2)_{1/6}$, which can couple to both up-type and down-type quarks, as the representative. For the electroweak triplet, we consider the representation $\mathbb{T}_3\equiv(3, 3)_{-1/3}$. 

\begin{figure}[!t]
\begin{center}
\hspace*{-0.0cm}
\includegraphics[width=0.48\textwidth]{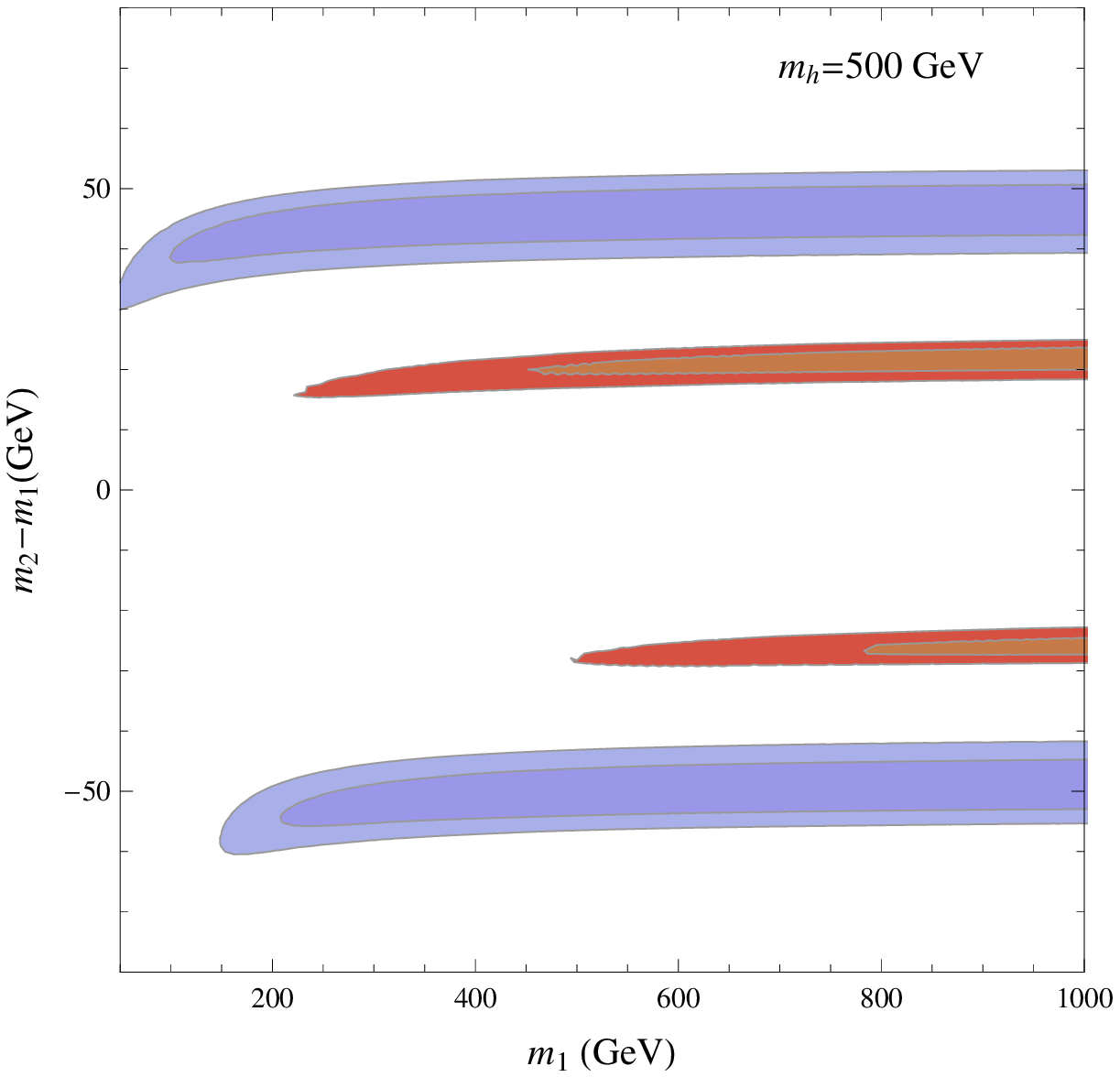} \hspace{3mm}
\includegraphics[width=0.48\textwidth]{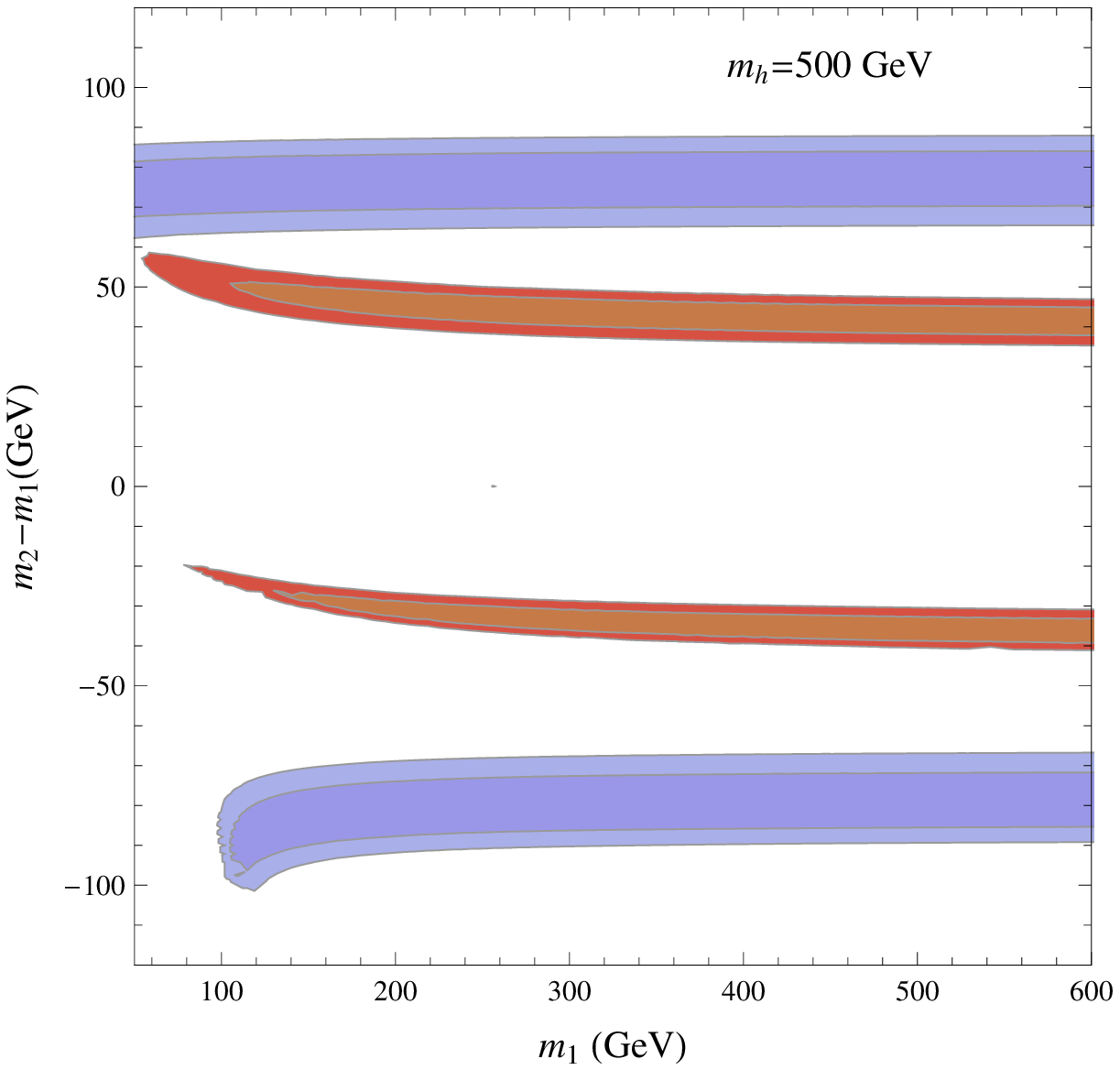}
\caption{Left panel: the allowed range of the mass split and the mass of octets for a 500 GeV mass Higgs from a fit to the $S$ and $T$ parameters. The two contours correspond to 68\% and 95\% C.L. The blue contours are for a $(8, 2)_{1/2}$ scalar, while the red for a $(8, 3)_1$ scalar. Right panel: the same as the left one but for $(3, 2)_{1/6}$ and $(3, 3)_{-1/3}$.}
\label{fig:colorscalarmass}
\end{center}
\end{figure}

We first consider the fit to the eletroweak precision observables. For electroweak doublets with the mass splitting $\delta\equiv m_2 - m_1 \ll m_1$, the modifications on $T$, $S$ and $U$ are approximately 
\beqa
\Delta T = \frac{d_c\,\delta^2}{12\,\pi\,s_W^2 M^2_W} \,, \qquad
\Delta S = - \frac{d_c\,Y\delta}{3\,\pi\,m_1} \,,\qquad
\Delta U = \frac{d_c\,\delta^2}{15\,\pi\,m_1^2} \,.
\eeqa
with $d_c=8$ for color octets and $d_c=3$ for color triplets. Similarly for electroweak triplets, we have 
\beqa
\Delta T = \frac{d_c\,\delta^2}{3\,\pi\,s_W^2 M^2_W} \,, \qquad
\Delta S = - \frac{4\,d_c\,Y\delta}{3\,\pi\,m_1} \,,  \qquad
\Delta U =  \frac{14\,d_c\,\delta^2}{15\,\pi\,m_1^2} \,.
\eeqa
Without  performing a numerical study, we can already know the constraints from $T$ and $S$ on the mass splitting and the lightest state mass. Since $\Delta T$ only depends on the mass splitting $\delta$, required values to fit the observed value of $T$ predict a constant value of $\delta$ that is independent of the overall mass. Once the $T$ parameter is satisfied, the constraints from $S$ can only impose a lower bound on the overall mass scale. The modification on the $U$ parameter has one more power of the heavy weak multiplet mass in the denominator than the $S$ parameter. This could be understood as when writing all new physics contributions as high dimensional operators in terms of the SM fields,  the $U$ parameter starts to get contribution from dimension six operators while the $S$ parameter is already modified by dimension five operators. 
The modifications of the $U$ parameter are numerically small and will be neglected in this section. Assuming $U=0$, we show the numerically fitted results in Fig.~\ref{fig:colorscalarmass} from just fitting the $S$ and $T$ parameters.  
One can see from those two plots that the allowed mass splittings are always below the $W$ gauge boson mass. The heavier state may decay into the light state plus an off-shell $W$, which will be discussed later for the collider phenomenologies.

\subsubsection{Modifications on the Higgs Boson Production}
\label{sec:higgsproduction}
In the SM, the Higgs boson is mainly produced from gluon fusion through the Higgs-gluon effective operator after integrating out the top quark
\beqa
{\cal L}^{eff} = - C_g \,\frac{h}{v}\,\frac{1}{4} G^a_{\mu\nu}G^{a\mu\nu}  \,.
\eeqa
If other heavy colored particles exist, they will also contribute to $C_g$. According to the low energy Higgs theorem, each new colored particle with mass $m(v)$ would contribute~\cite{Low:2009di}
\beq
\delta C_g = \delta b \frac{\alpha_s t_r}{4\pi v} \frac{\partial \log m(v)}{\partial \log v}\, ,
\eeq
where $\delta b$ is the particle's contribution to the $SU(3)_c$ gauge coupling $\beta$ function coefficient which equals 2/3 or 1/6 respectively for a Dirac fermion and a complex scalar and $t_r$ is the Dynkin index. The sign of $\delta C_g$ thus only depends on the $H$ dependence of the new colored state's mass. If $m(v)$ decreases with the Higgs VEV, $\delta C_g$ would be negative and the interference between SM and new particle would be destructive. 

In the presence of new colored scalars $\mathbb{O}$ or $\mathbb{T}$, one could write down in the Lagrangian the following quartic operators coupling $\mathbb{O}$, $\mathbb{T}$ to the Higgs, 
\beq
-\frac{\lambda_2}{2}\,H^\dagger H\,\mathbb{O}^\dagger \mathbb{O} \,, \quad\quad 
-\frac{\lambda_2}{2}\,H^\dagger H\,\mathbb{T}^\dagger \mathbb{T}  \,.
\eeq
After EWSB, the $\mathbb{O}(\mathbb{T})$ mass is then $m_1=\mu + \lambda_2 v^2/2$, where the constant $\mu$ is from the quadratic mass term $\mu \mathbb{O}^\dagger \mathbb{O}$ ($\mu \mathbb{T}^\dagger \mathbb{T}$). According to the argument at the beginning of this section, when the quartic coupling $\lambda_2 < 0$, one may have destructive interference and reduce the SM Higgs boson production cross section in the gluon fusion channel. Notice that $\mu$ should always be bigger than $|\lambda_2 v^2/2|$ to forbid a negative mass which will trigger the spontaneous breaking of $SU(3)_c$. Requiring the radiative corrections to $\lambda_2$ to be below 30\% of its tree-level value, the coupling is constrained to be $\lambda_{2} \lesssim 7$. 

\begin{figure}[!]
\begin{center}
\hspace*{-0.75cm}
\includegraphics[width=0.48\textwidth]{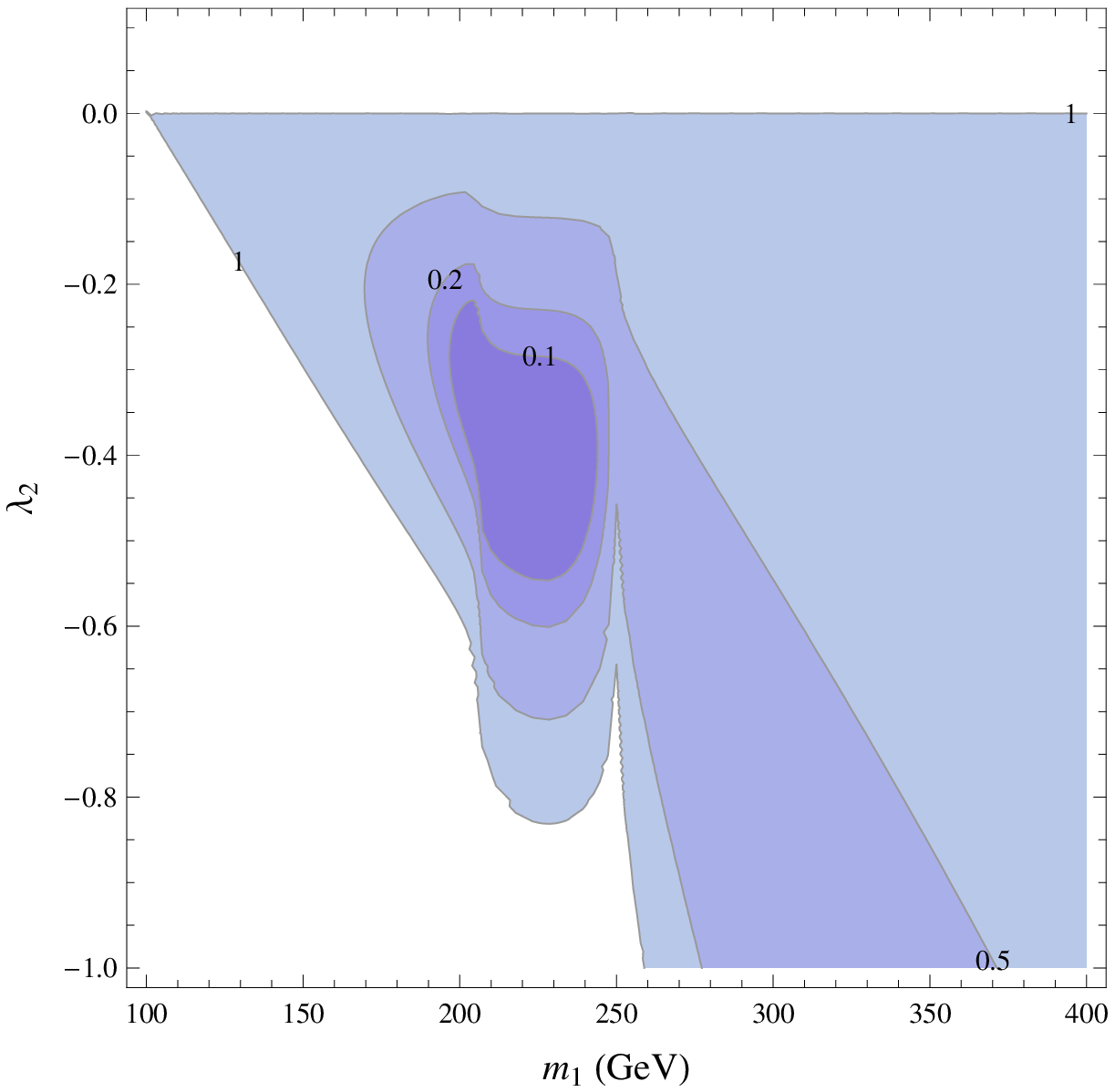} \hspace{3mm}
\includegraphics[width=0.48\textwidth]{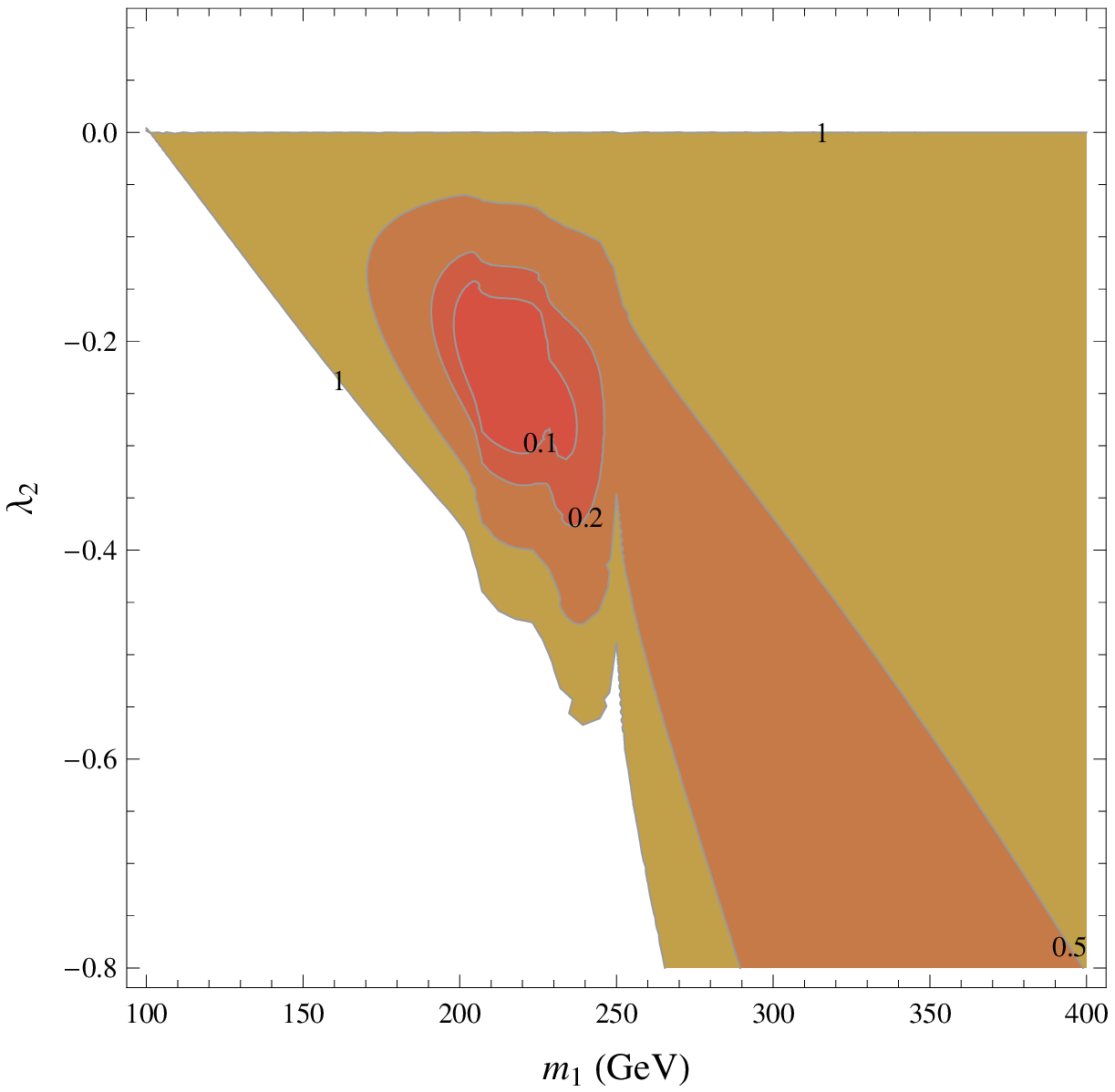}
\includegraphics[width=0.48\textwidth]{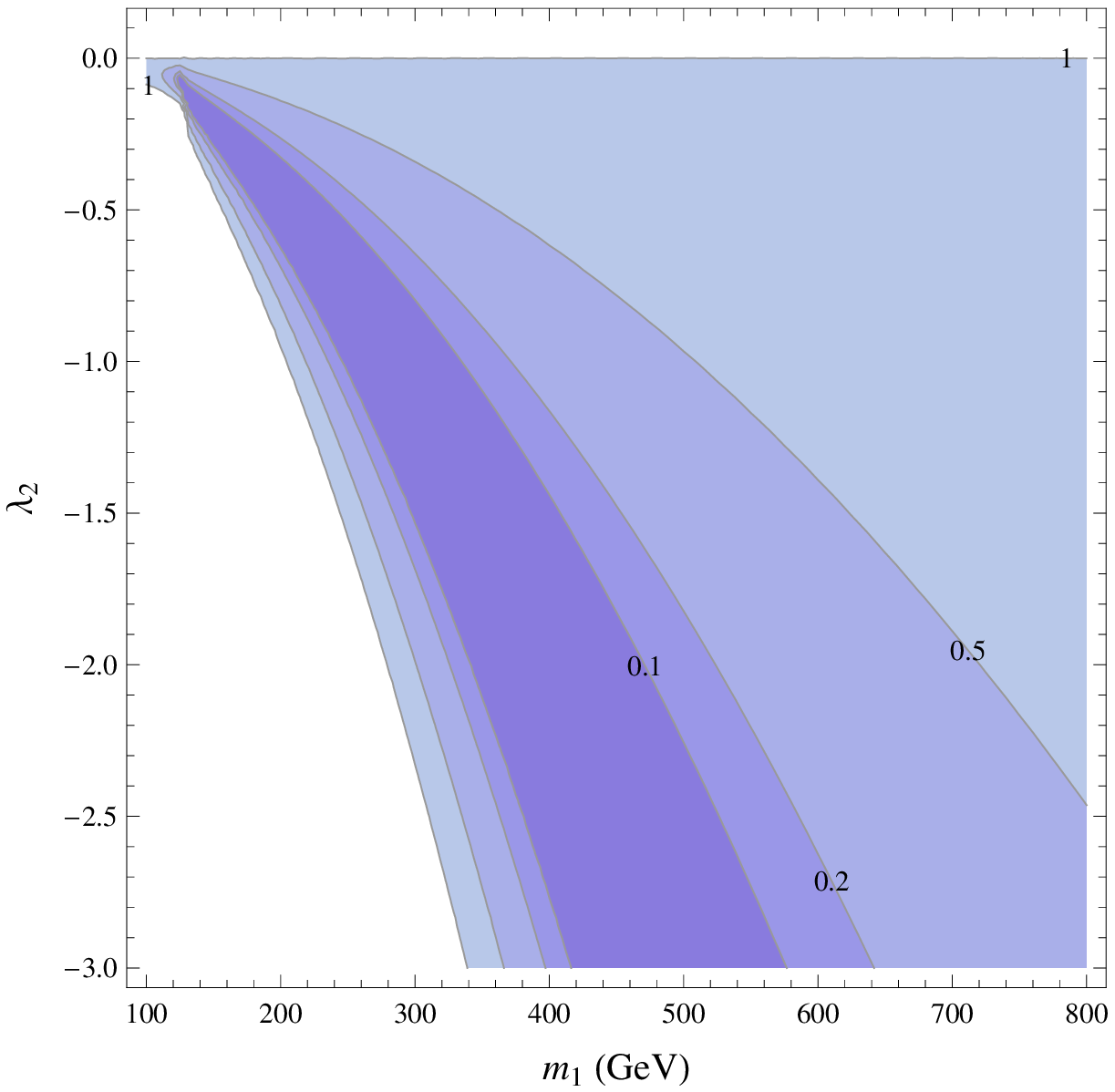} \hspace{3mm}
\includegraphics[width=0.48\textwidth]{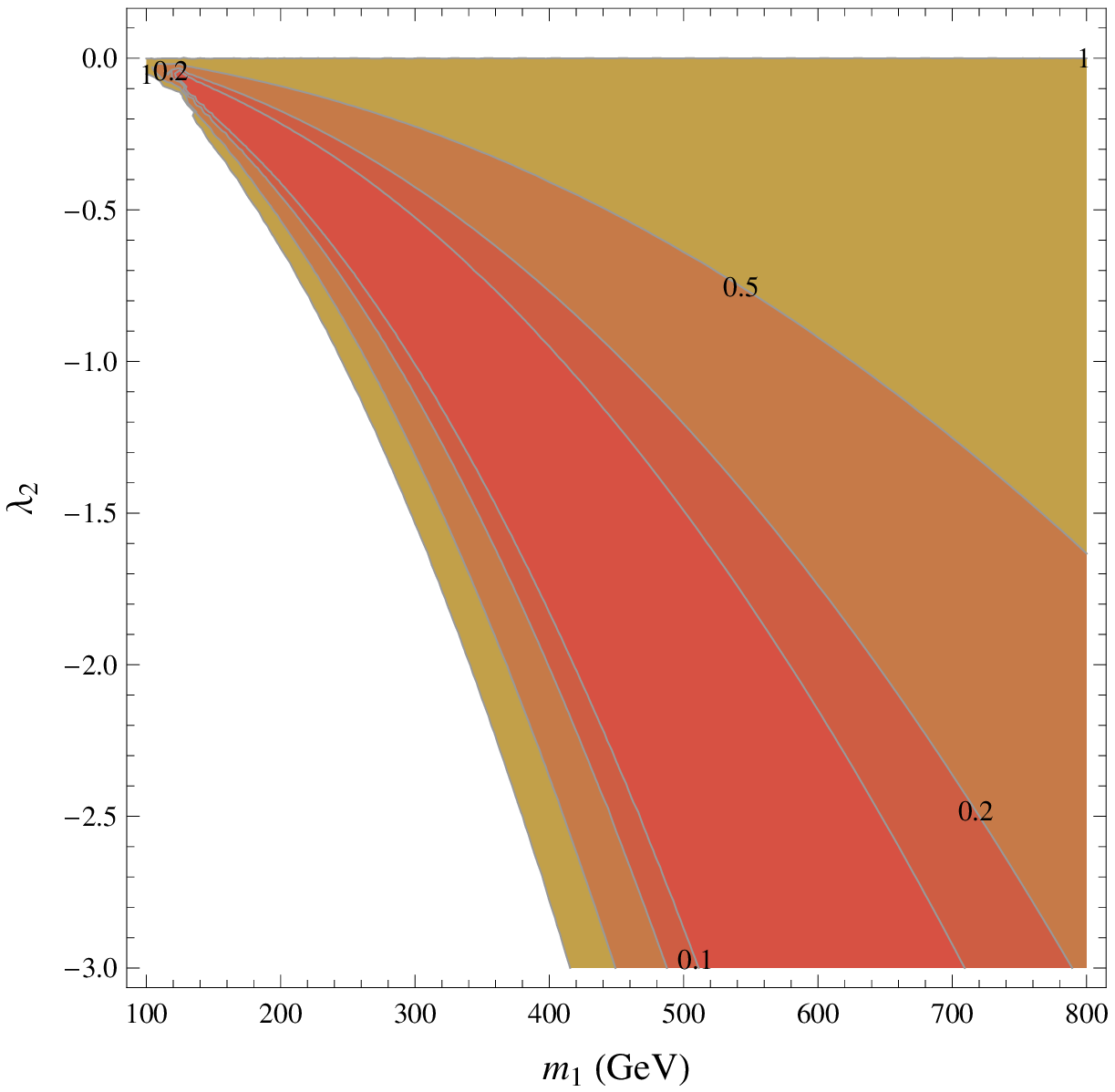}
\caption{Left panel: the ratio of production cross sections of a 500 GeV Higgs boson in the model with one additional color-octet weak doublet $(8, 2)_{1/2}$. Right panel: the same as the left one but for $(8,3)_{1}$. The two plots in the lower panels are for $m_h = 250$~GeV.}
\label{fig:productionRatioOctet}
\end{center}
\end{figure}

\begin{figure}[!]
\begin{center}
\hspace*{-0.75cm}
\includegraphics[width=0.48\textwidth]{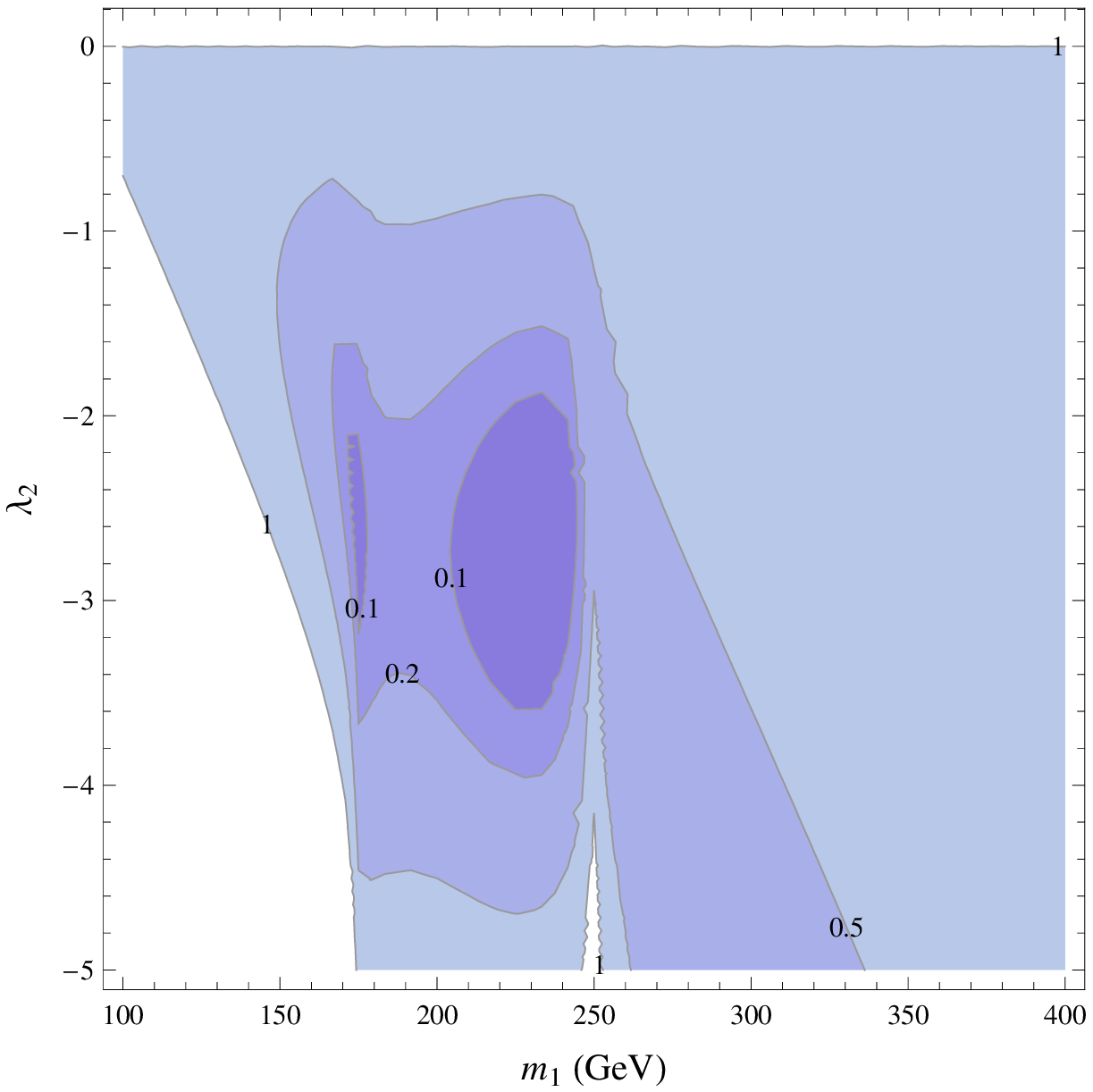} \hspace{3mm}
\includegraphics[width=0.48\textwidth]{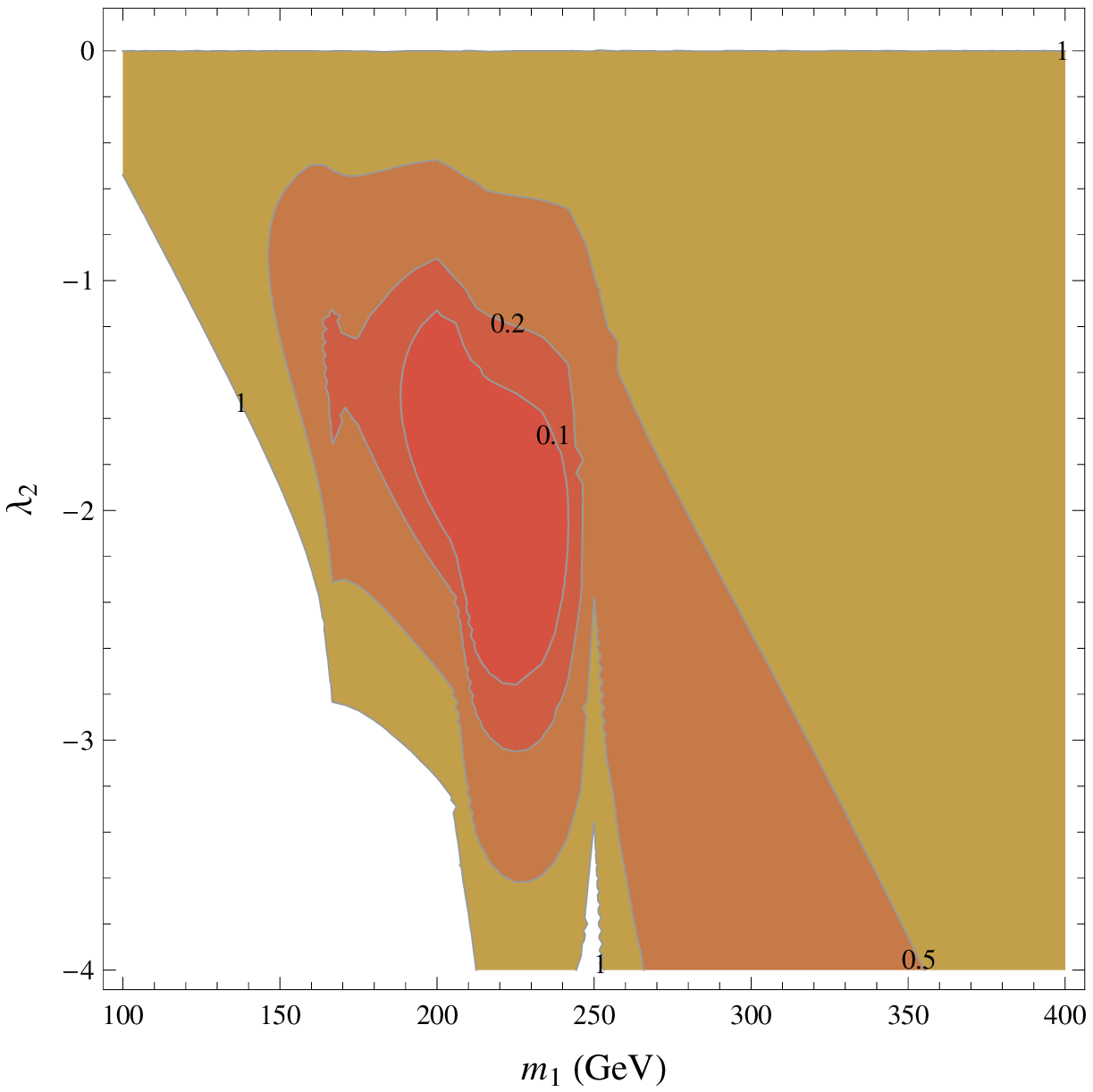} 
\includegraphics[width=0.48\textwidth]{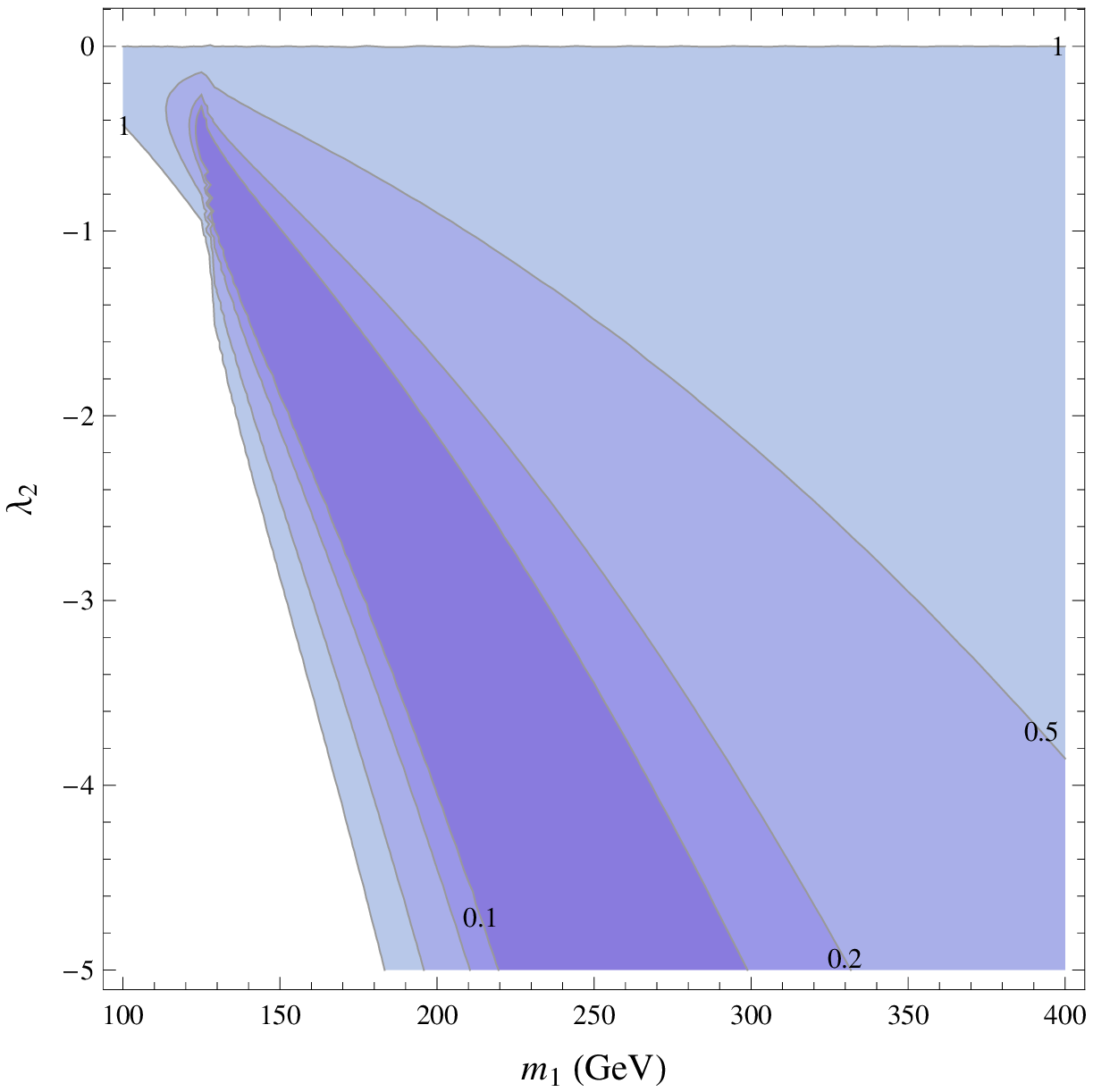} \hspace{3mm}
\includegraphics[width=0.48\textwidth]{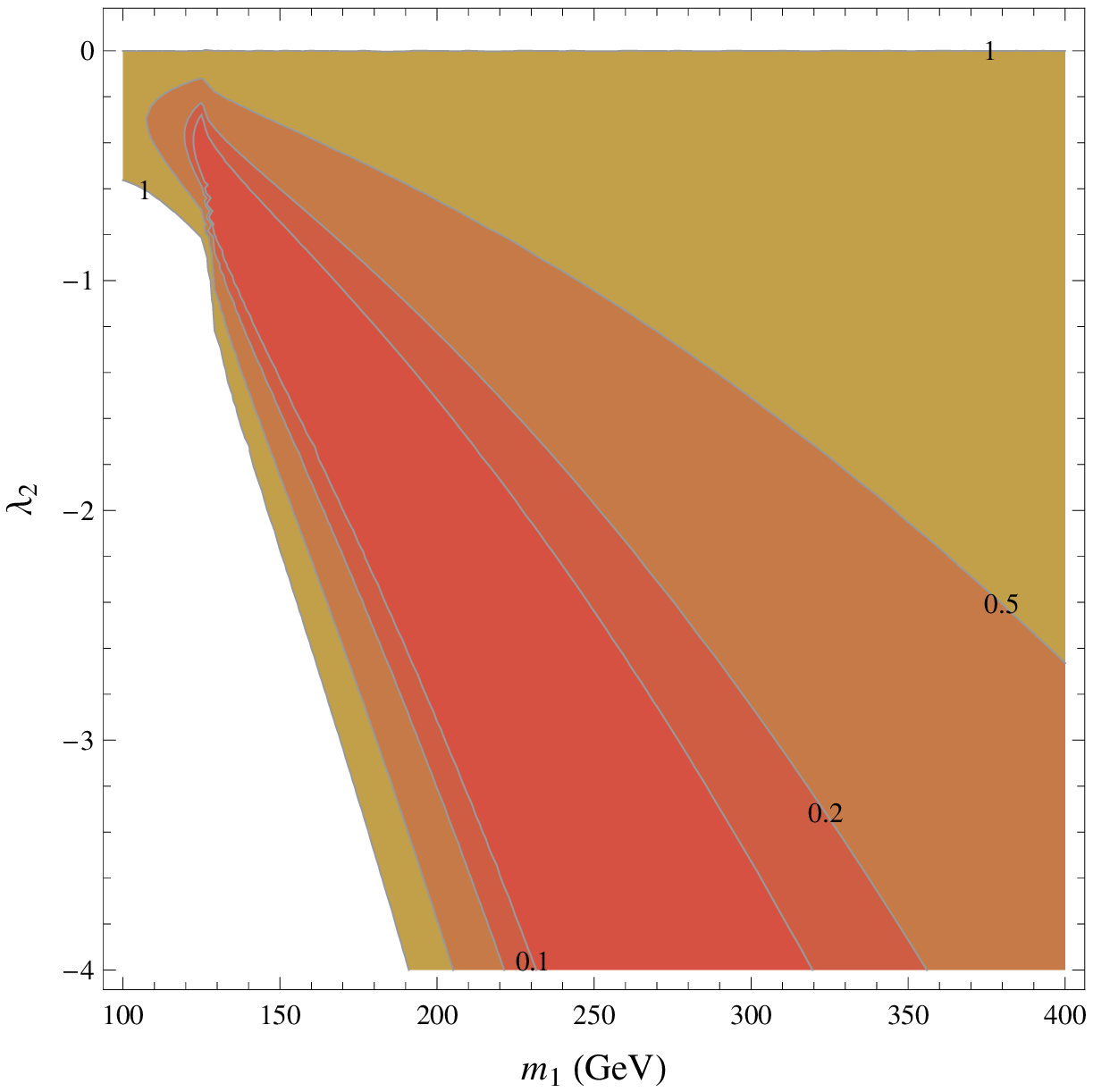}
\caption{The same as Fig.~\ref{fig:productionRatioOctet} but for color triplets $(3, 2)_{1/6}$ (left) and $(3, 3)_{-1/3}$ (right) with $m_h = 500$~GeV (upper) and $m_h = 250$~GeV (lower).}
\label{fig:productionRatioTriplet}
\end{center}
\end{figure}

So far the existing literature mainly focus on the enhancement of the Higgs production in the gluon fusion channel due to the colored scalars~\cite{Manohar:2006gz}\cite{Aglietti:2006tp}\cite{Bonciani:2007ex}\cite{Boughezal:2010ry}. However, we are interested in the destructive interference region where the coupling $C_g$ is reduced by the colored scalar loop (see Section~\ref{sec:HiggsEFT} for its formula). Note that when $m_h > 2m_t$, the SM contribution to $C_g$ contains both a real part and an imaginary part. To reduce the absolute value of $C_g$ by a certain amount, both parameters $\lambda_2$ and the mass $m_1$ have to be fixed. Only for a lighter Higgs boson with mass below both $2 m_t$ and $2 m_1$, $C_g$ is real and  the reduction of Higgs productions only constrain the ratio of $\lambda_2/m_1^2$ for a small splitting $\delta$ inside the scalar multiplet. In the limit that $m_h$ is much less than the masses of particles in the loop and neglecting $\delta$, the Higgs production cross section vanishes if the following relation between $m_1$ and $\lambda_2$ is satisfied
\beqa
\frac{m_1}{\sqrt{-\lambda_2}} = \frac{\sqrt{2n_{d_f}\,C(r)}}{4}\,v  \,,
\eeqa
where $v=\sqrt{2} v_{\rm EW} =246$~GeV; $n_{d_f}$ is the number of colored states and is 4 for a weak doublet and 6 for a complex weak triplet; $C(r)=\frac{1}{2}$ for fundamental representations and $3$ for adjoint representations. In Fig.~\ref{fig:productionRatioOctet} and Fig.~\ref{fig:productionRatioTriplet}, we show the ratios of the Higgs production cross section from gluon fusion in the SM plus a new colored state over that in the SM. 
 For a heavy Higgs boson with 500 GeV mass, the colored states are predicted to be between around 200 GeV to 250 GeV, if the SM Higgs boson production cross section is observed to be one tenth of the SM production cross section. For a lighter Higgs boson with  a 250 GeV mass, the new heavy colored state can be as heavy as 500 GeV.
 
\begin{figure}[!]
\begin{center}
\hspace*{-0.75cm}
\includegraphics[width=0.6\textwidth]{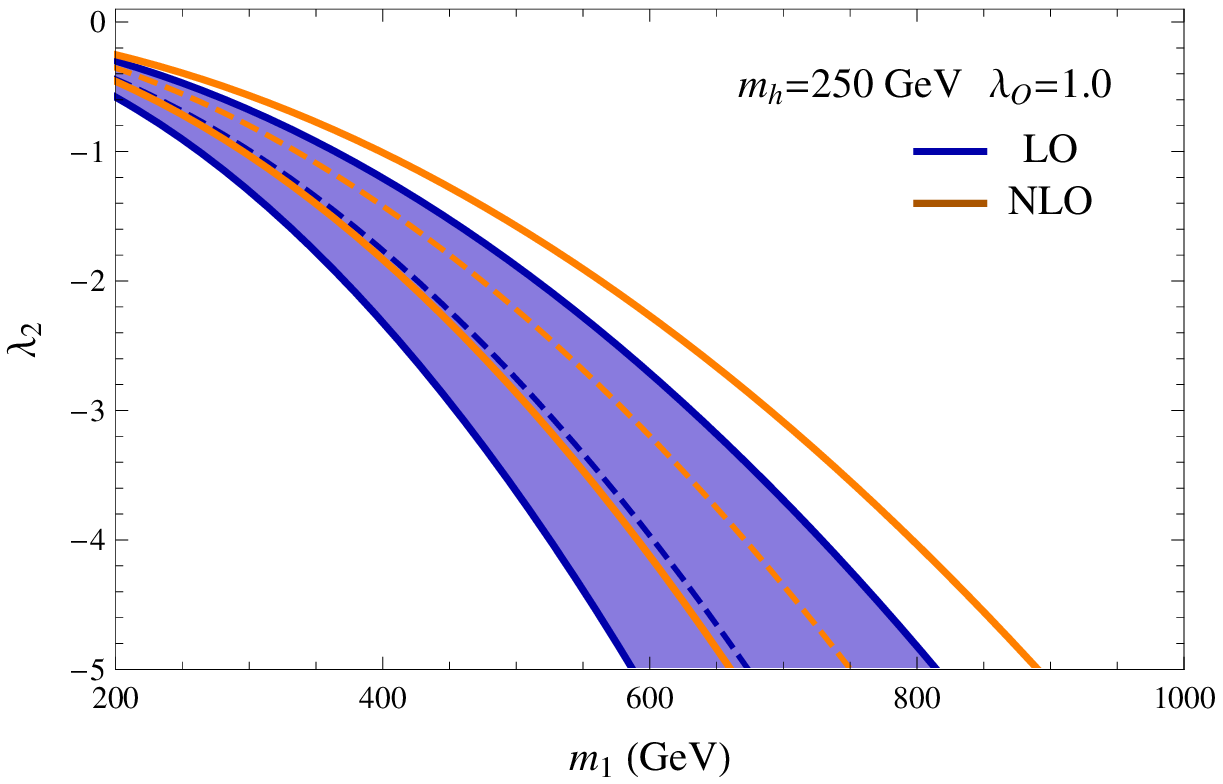} 
\caption{The region of parameter space in $\lambda_2$ and $m_1$ to reduce the $gg\rightarrow h$ production cross section as 10\% of the SM one. The new particle is a color-octet and weak-doublet scalar. The blue region is at the leading order in $\alpha_s$ and the orange region is at the next-to-leading order.}
\label{fig:ratioNLO}
\end{center}
\end{figure}

Since we have only considered here the reduction of Higgs production cross sections at the leading order in $\alpha_s$, one may worry about how stable this reduction is at the next-to-leading order (NLO)~\cite{Aglietti:2006tp}\cite{Bonciani:2007ex}\cite{Boughezal:2010ry}. Taking the heavy top quark and heavy colored-particle limit, using Eq.~(\ref{eq:NLOcg}) at NLO we present the modifications on the relation of $\lambda_2$ and $m_1$ for different $\lambda_{\mathbb{O}}$ (defined as the term $-g_s^2 \lambda_{\mathbb{O}} \mbox{Tr}\left[ S^2\right]^2$ in the Lagrangian) in Fig.~\ref{fig:ratioNLO} for $m_h = 250$~GeV, $\lambda_{\mathbb{O}} = 1.0$ and a color-octet weak-double scalar. 
Comparing the blue and orange regions in Fig.~\ref{fig:ratioNLO}, we have found the relation between $\lambda_2$ and $m_1$ is fairly stable to reduce the Higgs production cross section. For fixed $\lambda_2$, the relative modification on the new scalar mass is around 10\%.

\subsubsection{The properties of the colored particles}
\begin{figure}[!]
\begin{center}
\hspace*{-0.75cm}
\includegraphics[width=0.6\textwidth]{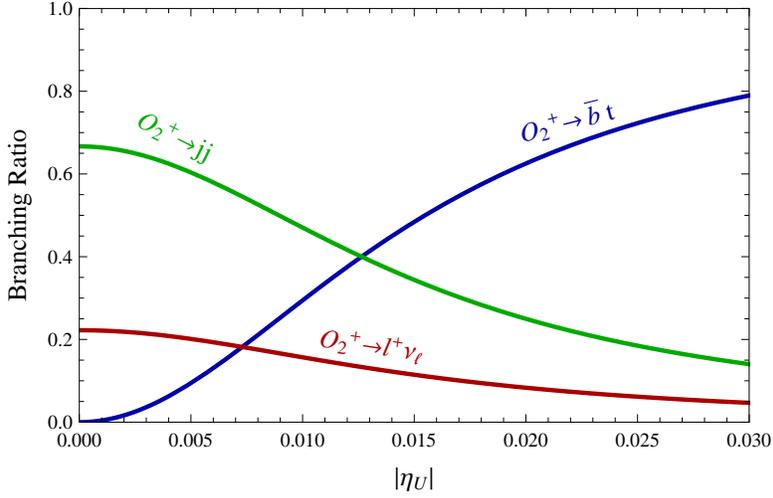}
\caption{The decaying branching ratios of the charged component of the color-octet and weak-doublet scalar. The heavier state mass is $m_2 = 264$~GeV and the lighter state mass is $m_1 = 220$~GeV.}
\label{fig:PBrOct}
\end{center}
\end{figure}

For the color octet and weak doublet particle $\mathbb{O}_2$, one can write down the Yukawa couplings to SM quarks at the renormalizable level. As discussed in Ref.~\cite{Manohar:2006gz}, the MFV assumptions can be realized for this particle and one can have the following interactions in the Lagrangian
\beqa
{\cal L} &=&  - \sqrt{2}\eta_U \bar{u}^i_R  \frac{m_U^i}{v}\,T^A\,u^i_L \mathbb{O}^{0A}_2 
+  \sqrt{2}\eta_U \bar{u}^i_R  \frac{m_U^i}{v}\,T^A\,V_{ij}\,d^j_L \mathbb{O}^{+A}_2 \nonumber \\
&& -\sqrt{2} \eta_{D}\bar{d}^i_R \frac{m^i_D}{v}\,T^A\,d^i_L \mathbb{O}_2^{0A\,\dagger} 
- \sqrt{2} \eta_{D} \bar{d}^i_R \frac{m^i_D}{v}\, V^\dagger_{ij} T^A u_L^j \mathbb{O}_2^{-A} \,+\, h.c. \,,
\eeqa
with $T^A$ as the $SU(3)_c$ generators and $m_{U\, D}^i$ are up-type (down-type) quark masses. Since the third-generations have the largest Yukawa couplings, the color octets prefer to decay into $t-$ or $b-$ quarks. The decays widths are 
\beqa
\Gamma(\mathbb{O}^{+}_2 \rightarrow t \bar{b} ) &=& \frac{|\eta_U|^2}{16\pi m_2^3} \left( \frac{m_t}{v} \right)^2 
|V_{tb}|^2 (m_2^2 - m_t^2)^2  \,, \nonumber \\
\Gamma(\mathbb{O}^{0}_{2\,R,I} \rightarrow b \bar{b} ) &=& \frac{|\eta_U|^2\,m_1}{16\pi} \left( \frac{m_b}{v} \right)^2  \,,
\nonumber \\
\Gamma(\mathbb{O}^{0}_{2\,R} \rightarrow t \bar{t} ) &=& \frac{m_1}{16\pi} \left( \frac{m_t}{v} \right)^2  
\left[  |{\rm{Re}}\,\eta_U|^2 \left( 1 - \frac{4 m^2_t}{m_1^2} \right)^{3/2} +  |{\rm{Im}}\,\eta_U|^2 \left( 1 - \frac{4 m^2_t}{m_1^2} \right)^{1/2} \right] \,, \nonumber \\
\Gamma(\mathbb{O}^{0}_{2\,I} \rightarrow t \bar{t} ) &=& \frac{m_1}{16\pi} \left( \frac{m_t}{v} \right)^2  
\left[  |{\rm{Re}}\,\eta_U|^2 \left( 1 - \frac{4 m^2_t}{m_1^2} \right)^{1/2} +  |{\rm{Im}}\,\eta_U|^2 \left( 1 - \frac{4 m^2_t}{m_1^2} \right)^{3/2} \right] \,, 
\eeqa
Through the weak interaction, the charged state can also decay into the neutral state plus a $W$ gauge boson. From Fig.~\ref{fig:colorscalarmass}, the mass splitting between those two states are below the $W$ gauge boson mass, so only off-shell decays are allowed and the decay width is (our result is different from Ref.~\cite{Djouadi:1995gv})
\beqa
\Gamma(\mathbb{O}^{+}_2 \rightarrow \mathbb{O}^{0}_{2\,I, R} + W^{+\,*} \rightarrow  \mathbb{O}^{0}_{2\,I, R} + \bar{f}f ) = \frac{a\,g_2^4}{128\,\pi^3}\,m_2\, G\left(\frac{m^2_1}{m^2_2}, \frac{M^2_W}{m^2_2}\right) \,,
\eeqa
with $a=\frac{1}{4} (\frac{1}{2})$ for a weak doublet (triplet) and with the function $G(k_i, k_j)$ as
\beqa
G(k_i, k_j) &=&\frac{1}{12 k_j} \left\{ 6(1 +k_i - k_j) k_j \sqrt{\lambda_{ij}} \left[ \arctan \left( \frac{k_i + k_j -1 }{\sqrt{\lambda_{ij} }}\right) + \arctan \left( \frac{k_i - k_j -1 }{\sqrt{\lambda_{ij} }} \right) \right] \right.\nonumber \\
&& \left. \hspace{-1cm}  -\, 3\left[ 1+(k_i - k_j)^2 - 2 k_j \right] k_j \log(k_i) \,+\,
(1- k_i) \left[ -2 (k_i -1)^2 + 9 (1 + k_i) k_j - 6 k_j^2 \right]
 \right\} 
\eeqa
and $\lambda_{ij}  = -1 + 2k_i +2 k_j -(k_i - k_j)^2$. Here, $f$ stands for SM fermions. We note that when $|m_1 - m_2| \ll m_1$, the above formula can be approximated in terms of the mass splitting $\delta= m_2 - m_1$ as
\beqa
\Gamma(\mathbb{O}^{+}_2 \rightarrow \mathbb{O}^{0}_{2\,I, R} + W^{+\,*} \rightarrow  \mathbb{O}^{0}_{2\,I, R} + \bar{f}f ) =  \frac{a\,\,g_2^4\, \delta^5}{240\,\pi^3\,M_W^4}  \,,
\eeqa
which agrees with the formula in Ref.~\cite{Dobrescu:2011px}. For $m_1 = 220$~GeV and $m_2 = 264$~GeV, we show the branching ratios of different decay channels in Fig.~\ref{fig:PBrOct}.

Beyond the renormalizable operators, one can also write down higher-dimensional operators to mediate the decays of colored states. For example, at dimension-6 level one can have
\beqa
\frac{d^{ABC}\,H^\dagger\,\mathbb{O}^A_2\,G^B_{\mu\nu}\,G^{C\,\mu\nu}}{M^2}\,, \quad
\frac{H^\dagger\,\mathbb{O}^A_2\,G^A_{\mu\nu}\,B^{\mu\nu}}{M^2}\,, \quad
\frac{\left[H^\dagger\tau^a\mathbb{O}^A_2\right]G^A_{\mu\nu}\,W^{a\,\mu\nu}}{M^2} \,,
\eeqa
which will mediate additional decay channels of the colored states as
\beqa
\mathbb{O}^+_2 \rightarrow g + W^+ \,,\quad 
\mathbb{O}^0_2 \rightarrow g + g \,, \quad
\mathbb{O}^0_2 \rightarrow g + \gamma / Z \,.
\eeqa

For the color-octet and weak triplet $\mathbb{O}_3$,  the analysis is very similar to the $\mathbb{O}_2$ except that we now have one more double-charged state $\mathbb{O}_3^{++}$, which will decay into the state $\mathbb{O}_3^{+}$ by an off-shell $W$ gauge boson. The following dimension-5 operators can couple this particle to the SM fermions
\beqa
\frac{\overline{Q}_L \mathbb{O}_3^{A,a\,*} T^A \tau^a H\, u_R }{M} \,, \quad
\frac{\overline{Q}_L \mathbb{O}_3^{A,a} T^A \tau^a \tilde{H}\, d_R }{M} \,,
\eeqa
where we have neglected the flavor indexes. 

For the color-triplet states, the higher isospine states will always decay into the lower one plus an off-shell $W$. Because of the specific hypercharge assignments, the lightest state must decay into at least one quark. The following operators can mediate its decays
\beqa
\frac{\epsilon^{ijk} H^\dagger \mathbb{T}_2^i u_R^{j\,T} {\cal C}  d_R^{k}}{M} \,,\quad 
\frac{\epsilon^{ijk} H \mathbb{T}_2^i d_R^{j\,T} {\cal C}  d_R^{k}}{M} \,,\quad 
\frac{\epsilon^{ijk} H^\dagger \mathbb{T}_3^i Q_L^{j\,T}\gamma^0 u_R^{k} }{M}  \,, \quad
\frac{\epsilon^{ijk} H \mathbb{T}_3^i Q_L^{j\,T}\gamma^0 d_R^{k} }{M}  \,.
\eeqa
Here, $\epsilon^{ijk}$ is a full anti-symmetric tensor in terms of QCD indexes with $i, j, k = 1, 2, 3$. Therefore, the lightest states of $\mathbb{T}_2$ and $\mathbb{T}_3$ should decay into different flavors of quarks. For sufficiently high values of $M$, we have the dominant decay channel of color triplets as
\beqa
\mathbb{T}_2^{2/3} &\rightarrow&  \mathbb{T}_2^{-1/3} + W^{+\,*}   \,, \quad\quad \mathbb{T}_2^{-1/3}  \rightarrow j + j  \,,\nonumber \\
\mathbb{T}_3^{2/3} &\rightarrow&  \mathbb{T}_3^{-1/3} + W^{+\,*}  \,,\quad
\mathbb{T}_3^{-1/3} \rightarrow  \mathbb{T}_3^{-4/3} + W^{+\,*} \,,\quad 
\mathbb{T}_3^{-4/3}  \rightarrow j + j \,.
\eeqa
%

\subsubsection{The collider constraints and signatures of colored states}
\begin{figure}[t!]
\begin{center}
\hspace*{-0.75cm}
\includegraphics[width=0.6\textwidth]{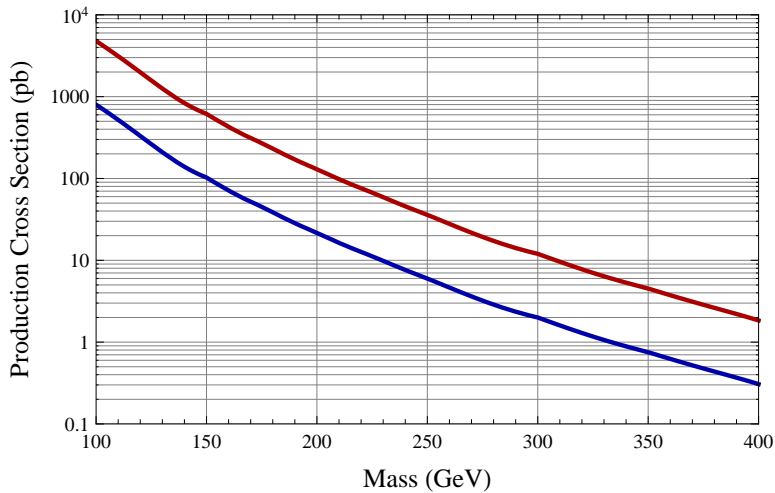}
\caption{The production cross sections of a complex color-octet (red and upper) and a complex color-triplet (blue and lower) at the leading order in QCD.}
\label{fig:PQCDprod}
\end{center}
\end{figure}

To reduce the Higgs boson production in the gluon fusion channel, the new colored states are predicted to have a mass from 200~GeV to 300~GeV, as can be seen from Fig.~\ref{fig:productionRatioOctet} and Fig.~\ref{fig:productionRatioTriplet}. Before we talk about the collider signatures of these new colored particles, we first calculate their production cross sections at the 7 TeV LHC. We use the FeynRules~\cite{Christensen:2008py} to generate our new physics models and use the MadGraph 5~\cite{Alwall:2011uj} to calculate the production cross sections. In Fig.~\ref{fig:PQCDprod}, we show the production cross sections for a complex color-octet (red) and complex color-triplet (blue) at the leading order from QCD interaction. For the same mass, the production cross section of the color-octet scalar is 6 times the color-triplet one, which is the ratio of the Dynkin indexes. 

With such large production cross sections for those colored particles, the LHC has a very good chance to discover them. Depending on different decay channels of those particles, we anticipate the following particles in the final state:
\beqa
pp &\rightarrow& \mathbb{O}^0_2 \mathbb{O}^{0\,*}_2 \rightarrow 4b\,, 4g\,, 2g+2\gamma\,, 2g+2 Z\,, 
\nonumber \\
pp &\rightarrow & \mathbb{O}^{+}_2 \mathbb{O}^{-}_2 \rightarrow W^{+\,*}+W^{-\,*} + (4b\,, 4g\,, 2g+2\gamma\,, 2g+2 Z) \,, \nonumber \\
pp &\rightarrow & \mathbb{O}^{++}_3 \mathbb{O}^{--}_3 \rightarrow 2W^{+\,*}+2W^{-\,*} + (4b\,, 4g\,, 2g+2\gamma\,, 2g+2 Z) \,,
\eeqa
for the color-octet and
\beqa
pp &\rightarrow& \mathbb{T}^{-1/3}_2 \mathbb{T}^{+1/3}_2 \rightarrow 4j\,, 
\nonumber \\
pp &\rightarrow &  \mathbb{T}^{2/3}_2 \mathbb{T}^{-2/3}_2 \rightarrow W^{+\,*}+W^{-\,*} + 4j \,, \nonumber \\
pp &\rightarrow &  \mathbb{T}^{2/3}_3 \mathbb{T}^{-2/3}_3 \rightarrow 2W^{+\,*}+ 2W^{-\,*} + 4j \,.
\eeqa

For $2g+2\gamma$ and $2g + 2Z$, there are no current experimental searches at the LHC, but one can refer to Ref.~\cite{Bai:2010mn} for existing phenomenological studies. For the $4b$ final state, the existing search for the MSSM Higgs at the CDF of Tevatron with 2.6~fb$^{-1}$~\cite{Aaltonen:2011nh}  does not constrain color-octets or color-triplets even when they have 100\% branching ratio to $2b$'s, while a new search aiming at the pair-produced colored states may constrain color-octets as pointed out in Ref.~\cite{Bai:2010dj}. For the four-jet final state, the recent searches at the ATLAS detector with 34~fb$^{-1}$ constrain the mass of a complex color-octet to be above 185~GeV (or below 100 GeV)~\cite{Aad:2011yh}. Assuming the same acceptance for the color-triplet, we found that the current four-jet analysis does not constrain the complex color-triplet scalars. One might hope that with more luminosity some constraints can be obtained in the near future if the low-$p_T$ jet triggers could still be used.

For the decay products of higher isospin states, we have additional off-shell $W^*$ gauge bosons. Depending on the $W^*$ gauge boson decaying channels, we may have multi-leptons plus multi-jets in the final state. Starting with the weak-triplet states ${\mathbb{O}}^{++}_3$ and $\mathbb{T}^{2/3}_3$, we can have two leptonic decays for two same-sign $W^*$'s and have same-sign dileptons plus jets and  missing energy in the final state. Based on 0.98 fb$^{-1}$ luminosity data, the CMS Collaboration has set limits on new physics production cross sections from three baseline selections. Considering that fact the leptons from the $W^*$ have smaller $p_T$, we take limits from the ``inclusive dileptons" baseline selection in Ref.~\cite{CMSdilepton} to set limits on the new particles considered here. We summarize the cuts used in their analysis:
\beqa
&& p_T(\mbox{jet}) > 40~\mbox{GeV}\; \mbox{with}\;|\eta|< 2.5; \mbox{at least two jets};  \nonumber \\
&& \mbox{two same-sign leptons with }p_T(\mbox{muon}) > 5~\mbox{GeV and } p_T(\mbox{electron}) > 10~\mbox{GeV}; \nonumber \\
&& H_T \equiv \sum p_T(\mbox{jet}) >  400~\mbox{GeV};\quad \missET > 50~\mbox{GeV} \,.
\eeqa
Using Pythia for showering and PGS with the CMS card for a detector simulation, we estimate the signal acceptance efficiency after passing those cuts. For the color-octet states with masses $(m_1=310, m_2 = 330, m_3 =350)$~GeV, we found that the signal acceptance passing those cuts are around 6\%. So, the signal will predict around 12 events after those cuts. The exclusion limit from CMS~Ref.~\cite{CMSdilepton} is 8.9 events at 95\% C.L. Therefore, the lower mass constraints on the lightest state for the color-octet weak-triplet scalar should be around 310 GeV, which is insufficient to hide a heavy Higgs boson as can be seen from Fig.~\ref{fig:productionRatioOctet}. A similar conclusion can be obtained for the color-triplet weak-triplet scalars. Although they have a smaller production cross section, their acceptance efficiency is higher than the color-octet weak-triplet scalar because of a larger mass splitting among different states. 

For the weak-doublet colored states, the pair-productions of $\mathbb{O}_2^+ + \mathbb{O}_2^-$ or $\mathbb{T}^{2/3}_2 + \mathbb{T}^{-2/3}_2$ generate opposite-sign dileptons plus multijets in the final state. With an integrated luminosity of 0.98 fb$^{-1}$, the current searches at CMS constrain our signal production cross sections. From Ref.~\cite{CMSdilepton2}, we summarize their cuts below:
\beqa
&& \mbox{at least two leptons with } p_T >10~\mbox{GeV}, |\eta|< 2.5~\mbox{for } e^\pm~(2.4~\mbox{for } \mu^\pm)\,; \nonumber \\
&& \mbox{the leading lepton having } p_T > 20~\mbox{GeV}\,; \nonumber \\
&& \mbox{remove the regions } 76~\mbox{GeV} < m_{\ell^+ \ell^-} < 106~\mbox{GeV}\; \mbox{and} \;
      m_{\ell^+ \ell^-}  < 12~\mbox{GeV}\,; \nonumber \\
&& \mbox{at least two jets (anti-$k_T$ clustering) with } p_T > 30~\mbox{GeV and } |\eta| < 3.0\,; \nonumber \\
&&  \mbox{or high $H_T$ signal region}: H_T  >  600~\mbox{GeV};\quad \missET > 200~\mbox{GeV} \,.
\eeqa
 After PGS detector simulations, we found that the acceptance efficiency for the color-octet $\mathbb{O}_2$ with $m_1=220$~GeV and $m_2=264$~GeV  is  around 0.01\% (the high $\missET$ cuts in Ref.~\cite{CMSdilepton2} provide an even smaller efficiency).  Multiplying the leptonic decaying branching ratio, we found less than one signal events after cuts.  However, if we don't impose the stringent $\missET$ or high $H_T$ and only require $\missET > 50$~GeV and $H_T > 100$~GeV, we find that the signal acceptance efficiency is around 6\% or 77 events at 0.98 fb$^{-1}$. The observed number of events in Ref.~\cite{CMSdilepton2} is 2481, which allows 98 signal events at 95\% C.L. if neglecting the systematic errors. So, this color-octet weak-doublet scalar with $(m_1, m_2) = (220, 264)$~GeV is very close to be ruled out by the current searches optimized to SUSY models.
  
 Performing a similar analysis for the color-tripet weak-doublet scalar $\mathbb{T}_2$, we find that the lower cuts with $\missET > 50$~GeV and $H_T > 100$~GeV provide a better constraint. For $(m_1, m_2) = (220, 295)$~GeV, the acceptance efficiency is around 15\%. Using the production cross section in Fig.~\ref{fig:PQCDprod} and multiplying the leptonic branching ratio, there are 16 signal events after cuts, which is below the current experimental error bars. So, this color-triplet weak-doublet scaler is not constrained from the opposite-sign dilepton searches. 
  
To summarize, a color-triplet weak-doublet scalar $\mathbb{T}_2$ with a mass splitting around 75 GeV between its two isospin states and a 220 GeV mass for the $I_3=-1/2$ state can hide a heavy Higgs boson with $500$~GeV. The other three options, $\mathbb{T}_3$, $\mathbb{O}_2$ and $\mathbb{O}_3$ are in conflict with the current same-sign or opposite-sign dilepton searches.  For a lighter Higgs mass around $250$~GeV, the constraints become weaker because of a smaller mass splitting $\delta$ is required to fix the electroweak observables. Since reducing the Higgs production cross section in gluons fusion channel only constraints the ratio of $m_1$ and $\sqrt{-\lambda_2}$, a heavier colored state with suppressed paired productions at the 7 TeV LHC is always allowed from the current searches.  Interestingly, a large value of $|\lambda_2|$ is required from Fig.~\ref{fig:productionRatioOctet}  and  Fig.~\ref{fig:productionRatioTriplet} for a heavier $m_1$. It might be interesting to look for the associated Higgs productions at the 14 TeV LHC as $p p \rightarrow \mathbb{T} \mathbb{T}^* h~\mbox{or}~\mathbb{O} \mathbb{O}^* h$.

\subsection{Fermions}
\label{sec:QCDfermions}
Additional vector-like fermions mixing with top-quark or bottom-quark can also modify electroweak observables~\cite{Peskin:2001rw}. One possibility is to introduce a vector-like $SU(3)_c$ triplet $\chi$, like in the ``top-color seesaw" model by Dobrescu and Hill~\cite{Dobrescu:1997nm}. Specifically, one can have the quantum numbers of $\chi$ to be the same as the right-handed top quark $(3, 1)_{2/3}$. The relevant parts in the Lagrangian are
\beqa
\lambda_1 \tilde{H} \overline{Q}_L t_R \,+\, \lambda_2\,\tilde{H} \overline{Q}_L \chi_R \,+\, \mu\,\overline{\chi}_L \chi_R \,,
\label{eq:vectorfermion}
\eeqa
where $Q_L = (t_L, b_L)^T$. Diagonalizing the mass matrix and under the assumption $\lambda_2 v_{\rm EW} \ll \mu$, we have the top quark mass to be $m_t \approx \lambda_1 v_{\rm EW}$ and the top-partner to be $m_{t^\prime} \approx \mu$. The left-handed mixing angle is 
\beqa
\beta_{L} &=& \frac{1}{2}\,\arctan{\frac{2 \lambda_2\, v_{\rm EW}\, \mu}{\mu^2 - (\lambda_1^2 + \lambda_2^2) v_{\rm EW}^2  }} \approx \frac{\lambda_2 \, v_{\rm EW}}{\mu} \,,
\label{Eq:LHmixing}
\eeqa
in the limit $\lambda_2 v_{\rm EW} \ll \mu$.

\begin{figure}[!h]
\begin{center}
\hspace*{-0.75cm}
\includegraphics[width=0.48\textwidth]{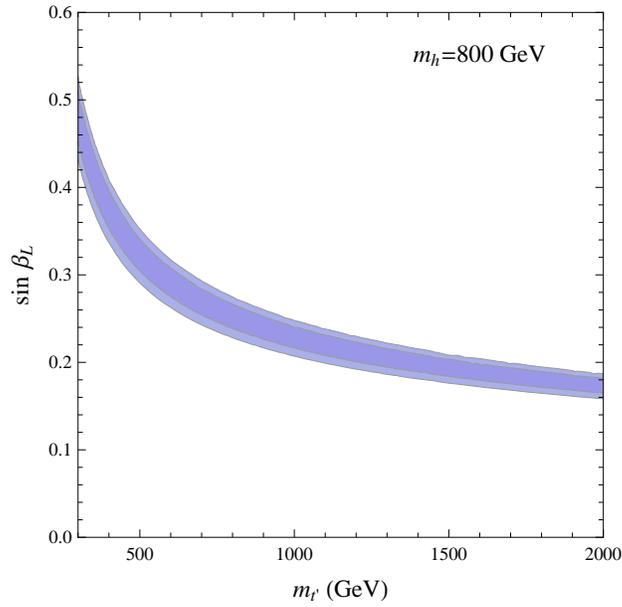} 
\caption{The allowed range of the $t^\prime$ mass and the left-handed mixing angle $\beta_L$ for a 800 GeV mass Higgs from a fit to the $S$ and $T$ parameters. The two contours correspond to 68\% and 95\% C.L.}
\label{fig:Ptprime}
\end{center}
\end{figure}

The contribution due to
the fermion loop is~\cite{Lavoura:1992np, He:2001fz, Bai:2008gm}  
\beqa
\Delta T_{f}&=& \frac{3\,s_{\beta_{L}}^{2}}{16 \pi 
\sin^{2}{\theta_{W}}\cos^{2}{\theta_{W}}}\left[
W_{1}(y_{\chi},y_{b})\,-\,W_{1}(y_{t},y_{b})\,-\,
c_{L}^{2}\,W_{1}(y_{t},y_{\chi})\right]~,  
\nonumber \\ [0.5em]
\Delta S_{f}&=& \frac{3\,s_{\beta_{L}}^{2}}{2 \pi} \left[
W_{2}(y_{\chi},y_{b})\,-\,W_{2}(y_{t},y_{b})\,-\,
c_{L}^{2}\,W_{3}(y_{t},y_{\chi})\right]~,
\eeqa
where $y_{i}\equiv m^{2}_{i}/M^{2}_{Z}$, and $s_{\beta_{L}}$,
$c_{\beta_{L}}$ are short notations for $\sin{\beta_{L}}$ and
$\cos{\beta_{L}}$. The approximation is based on $\mu \gg \lambda_i \, v_{\rm EW}$.  Here the
functions $W_{1}$, $W_{2}$ and $W_{3}$ are defined in Appendix~\ref{sec:Wfunction}. In terms of the mixing angle and the $t^\prime$ mass, we show the parameter space to fit $T$ and $S$ parameters with a very heavy $m_h=800$~GeV in Fig.~\ref{fig:Ptprime}.

The relevant couplings for collider phenomenologies of $t^\prime$ are
\beqa
{\cal L} \supset \frac{e\,s_{\beta_L}\,c_{\beta_L} }{ 2 s_W c_W} Z_\mu \left( \bar{t^\prime}_L \gamma^\mu t_L   + \bar{t}_L \gamma^\mu t^\prime_L \right) + \frac{e\,s_{\beta_L}}{\sqrt{2} s_W} \left( \bar{t^\prime}_L \gamma^\mu b_L + \bar{b}_L \gamma^\mu t^\prime_L \right) \,,
\eeqa
where the coupling to the Higgs boson depends on the right-handed mixing angle and is not shown here. Other than pair-productions from QCD, this top-partner $t^\prime$ can also be produced singly together with $b$  or $t$ from weak gauge bosons. The main decay channels would be $t^\prime \to t + Z$ or $t^\prime \to b + W^+$. 

The existing searches for a vector-like top quark at D0~\cite{Abazov:2010ku} have considered single production of $t^\prime$ together with a light quark by including both $s$- and $t$- channels. Generically, the coupling of $t^\prime$ to light quarks should have additional CKM-suppressing factors.  So, we don't anticipate a stringent constraints on our model parameters space. At the LHC, a heavy $t^\prime$ can be produced from QCD pair-production. The final state could be $t\bar{t}+2Z$, $b\bar{b} + W^+ W^-$ or $t \bar{b} + Z + W^-$. The existing search at CMS using 1.14 fb$^{-1}$~\cite{Chatrchyan:2011ay} has considered the QCD pair-production of $t^\prime$ and sets a constraint on the mass of $t^\prime$ to be above 475 GeV at 95\% C.L., where 100\% branching ratio has been assumed for $t^\prime \to t + Z$. So, we can see that there is a long way to go to rule out the scenario with a new vector-like fermion.

According to the discussions in Sec~\ref{sec:higgsproduction}, we found that the vector-like fermion with the couplings in Eq.~\ref{eq:vectorfermion} does not contribute to $C_g$ and thus has no effect on the Higgs production cross section. As a result, it could not hide the Higgs though it could still remedy the electroweak precision problem of a Higgs with mass above the LHC reach or models with strong dynamics triggering EWSB. However, one might build more complex models similar to those little Higgs models where the fermion mass has a non-linear dependence on the Higgs VEV. It is shown in~\cite{Low:2009di} that additional fermions in those models give a negative contribution to $C_g$ and thus reduce the Higgs production rate in the gluon fusion channel.

\section{Additional Vector Gauge Boson $Z^\prime$}
\label{sec:zprime}
As pointed in the literature~\cite{Hewett:1988xc, Peskin:2001rw, Chanowitz:2008ix, Langacker:2008yv}, additional $Z^\prime$ gauge boson mixing with the SM $Z$ boson can provide a positive contribution to the $T$ parameter and hence accommodate a heavy Higgs boson. Following Ref.~\cite{Peskin:2001rw}, the $2\times 2$ mass mixing matrix can be expressed as
\beqa
{\cal M}^2 = 
\left(
\begin{array}{cc}
m^2 &  \gamma M_Z^2 \\
\gamma M_Z^2 & M^2 
\end{array}
\right) \,,
\label{eq:zprimemassmatrix}
\eeqa
where $\gamma$ is a model-dependent parameter of order 1. In the limit $M_Z \ll M$, the $Z^0$ boson mass is approximately $M_Z^2=m^2(1-\delta)$ and the physical $Z^0$ boson contains a small fraction of $Z^\prime$ with the mixing angle denoted as $\xi$, where
\beqa
\delta = \gamma^2\,\frac{M_Z^2}{M^2}\,, \qquad \xi = \gamma\,\frac{M_Z^2}{M^2} \,.
\eeqa
Only the $T$ parameter is modified by the $Z-Z^\prime$ mixing and has a simple relation to $\delta$
\beqa
\alpha\,T = \delta =  \gamma^2\,\frac{M_Z^2}{M^2}\,.
\eeqa
For a heavy Higgs boson with 800 GeV mass, a fit to the oblique parameters fixes
\beqa
\delta = (3.1\pm 0.3)\times 10^{-3} \,, \qquad \mbox{or}\qquad
\frac{\gamma M_Z}{M}  = 0.055\pm 0.002 \,.
\label{eq:zprimemixingvalue}
\eeqa
The modification of the weak mixing is $\Delta \sin^2 \theta_W = - s_W^2 c_W^2/(c_W^2 - s_W^2)\,\delta \approx -0.001$, which is below the current experimental error bar~\cite{Nakamura:2010zzi}, by neglecting the additional model dependent couplings of $Z^\prime$ to SM fermions. The modification of the leptonic decay width of $Z^0$, $\Delta \Gamma_l \approx 100\,\delta = 0.31$~MeV, is also below the experimental error~\cite{Nakamura:2010zzi}.  

The first class of models we want to consider is the sequential $Z^\prime$ model including Kaluza-Klein excitations of the SM gauge bosons~\cite{Rizzo:1999br}. Typically, there is also a $W^\prime$ gauge boson in the spectrum. So, the modification of the $W$ gauge boson mass is $c_W^2 M_W\delta /(2c_W^2 - 2s_W^2) \approx 0.18$~GeV, which is well above the current experimental error of the $W$ gauge boson mass $M_W = 80.399\pm 0.023$~GeV~\cite{Nakamura:2010zzi}. Therefore, we conclude that it is unlikely to have a sequential $Z^\prime$ to accommodate a very heavy Higgs boson. 

The second class of models is to consider the three generations of SM fermions charged under $U(1)_X$, which is a linear combination of $U(1)_Y$ and $U(1)_{B-L}$. For this class of models, the gauge anomalies are cancelled without including additional SM singlet fermions~\cite{Appelquist:2002mw}. Defining a mixing angle $\theta_X$, the charge of SM fermions under $U(1)_X$ is given by $Q_X = \cos{\theta_X} Y/2 + \sin{\theta_X} (B-L)/2$. Referring back to the notation in Eq.~(\ref{eq:zprimemassmatrix}), we have $\gamma = \cos{\theta_X}\,g_X/g_Z$ with $g_X$ as the gauge coupling of $U(1)_X$ and $g_Z = g_2/c_W$. Using the model-independent constraints from Tevatron (from dilepton final states) in Ref.~\cite{Carena:2004xs}, one can relate this model to their $U(1)_{q+xu}$ class of models by $\tan{\theta_X} = (4-x)/(x-1)$.  The constraints on the general mixing parameter $\delta$ is  $\delta < 1/(30.1 + 15.5 \tan{\theta_X})^2 < 1.1\times 10^{-3}$, valid for $0< \theta_X \le \pi/2$~\cite{Chanowitz:2008ix}. Compared to the required value for $\delta$ in Eq.~(\ref{eq:zprimemixingvalue}), this class of $Z^\prime$ models is unlikly to accommodate a heavy Higgs boson.

The third class of models is motived by the grand unified theories. Following Ref.~\cite{Peskin:2001rw}, one can define a mixing angle $\theta$ between two $U(1)$ gauge bosons such that $\theta=0$ one has the $SO(10)$ boson $\chi$ and $\theta=\pi/2$ one has the $E_6$ boson $\psi$. In general, there are two Higgs doublets developing VEVs with $\langle H_u \rangle /\langle H_d \rangle  =\tan{\beta}$. The parameter $\gamma$ in Eq.~(\ref{eq:zprimemassmatrix}) is given by
\beq
\gamma = 2 s_W \,\sin^2{\beta}\, \left( \cos{\theta} \frac{1}{\sqrt{6}} - \sin{\theta} \sqrt{\frac{5}{18}} \right) 
+ 2 s_W \,\cos^2{\beta}\, \left( \cos{\theta} \frac{1}{\sqrt{6}} + \sin{\theta} \sqrt{\frac{5}{18}} \right) \,,
\eeq
where the gauge coupling relations $g_X^2 = \frac{3}{8} g_Y^2$ for $U(1)_\chi$ and $g_X^2 = \frac{5}{8} g_Y^2$ for $U(1)_\psi$ have been used. The latest searches of $Z^\prime$ at ATLAS~\cite{Collaboration:2011dca} with around 1.1~fb$^{-1}$ luminosity constrain the mass of $Z_\chi$ above 1.64 TeV and the mass of $Z_\psi$ above 1.49 TeV  at 95\% C.L. This translates into a constraint on the mixing parameter $\delta$ as
\beqa
\delta < 0.5\times 10^{-3}\quad \mbox{     for } Z_\chi\,,     \qquad
\delta < (\sin^2{\beta} - \cos^2{\beta} )^2\, 1.0  \times 10^{-3}\quad  \mbox{      for }  Z_\psi\,.
\eeqa
So, we conclude that the $E_6$ motivated $Z^\prime$'s are also unlikely to make a heavy Higgs boson consistent with electroweak precision observables. 

Although the generation independent models have stringent constraints, other flavor-dependent $Z^\prime$~\cite{Harris:1999ya}, especially if it is leptophobic, can have a less stringent constraint. They may relax the electroweak precision constraints on a heavy Higgs boson and should be searched for at the LHC.

\section{Non-linearly Realized EWSB}
\label{sec:nonlinear}
So far we have discussed a heavy Higgs with mass below 600 GeV. A much heavier Higgs will evade the current Higgs searches at the LHC. However,  it will have a large width comparable to its mass, e.g., for $m_h =$ 1 TeV, its width is 667 GeV, which could no longer be treated as an elementary particle. Instead it is more appropriate to describe EWSB at low energy by a non-linear electroweak chiral Lagrangian with a $\Sigma$ field: $\Sigma=e^{i\pi^a\sigma^a/v}$~\cite{Longhitano:1980tm, Longhitano:1980iz, Appelquist:1993ka}. $\Sigma$ transforms linearly under the $SU(2)_L \times SU(2)_R$ as $\Sigma \to L\Sigma R^\dagger$. This low energy effective theory might be completed into a strongly-interacting UV theory such as technicolor models~\cite{Weinberg:1979bn, Susskind:1978ms, Holdom:1984sk, Appelquist:1986an} or 5D Higgsless models~\cite{Csaki:2003zu}. In this case, one could not probe EWSB directly by detecting a Higgs boson though there exist indirect collider signals. In this section, we will discuss one such collider signature predicted by fixing the EWPT in non-linearly realized EWSB. 

It is known that EWPT is a major stumbling block for scenarios breaking electroweak symmetry with strong dynamics~\cite{Peskin:1990zt, Peskin:1991sw}. Naive estimates for QCD-like theories indicate a positive order one $S$ parameter as at the perturbative level, $S$ counts the number of degrees of freedom participating the electroweak sector. One possible solution to this $S$ problem, which was explicitly realized in the Randall-Sundrum setup by choosing particular fermion bulk profiles~\cite{Cacciapaglia:2004jz, Cacciapaglia:2004rb}, is to obscure the oblique corrections by non-oblique corrections~\cite{Grojean:2006nn}. In the operator language, the new physics generates a particular combination of higher dimensional operators contributing to both oblique and non-oblique corrections, which is poorly constrained. More concretely, the operators contributing to $S, T, U$ are combined with operators coupling fermions to gauge bosons, which by equation of motion, are equivalent to operators modifying triple gauge boson couplings (TGC). For instance from Ref.~\cite{Grojean:2006nn}, 
\beqa
&&-16 \pi\,{\cal O}_S + g^{ 2}( {\cal O}_q^1+ {\cal O}_\ell^1) = {\cal O}_{3V}, \nonumber \\
{\rm with} \; &&{\cal O}_S= -\frac{1}{32\pi} g g^\prime B^{\mu\nu} {\rm Tr}(W_{\mu\nu} T)\,, \quad  {\cal O}_f^1=i \bar{f}_L\gamma^\mu V_\mu f_L\,, \quad {\cal O}_{3V}=i g\, {\rm Tr}\left(W^{\mu\nu}\left[V_\mu, V_\nu\right]\right),
\label{eq:tgc}
\eeqa
where  $f=q\,{\rm or}\,\ell$ and $g$ and $g^\prime$ are couplings of SM $SU(2)_w$ and $U(1)_Y$; $V_\mu=(D_\mu \Sigma)\Sigma^\dagger$ and $T=\Sigma \sigma^3\Sigma^\dagger$. Currently TGCs are best constrained by measurements at LEP~\cite{Delmeire:2005ir}, e.g., $W$-pair production $e^+e^- \to W^+W^-$, which bounds the coefficient of ${\cal O}_{3V}$ in Eq.~(\ref{eq:tgc}) to be smaller than 0.05.~\footnote{The constraints on the coefficient are obtained from the limits on the Hagiwara-Peccei-Zeppenfeld-Hikasa (HPZH) triple-gauge-vertex parameters~\cite{Hagiwara:1986vm} and the relation between the HPZH parameters and the chiral Lagrangian coefficients~\cite{Chivukula:2005ji}.}
Notice that such a mechanism could also be applied to models with an elementary Higgs~\cite{Carena:2003fx, Agashe:2003zs}. It means that a better measurement of the gauge boson self couplings could be very useful for constraining new physics.

Future experiments may improve the limits on TGCs. An analysis of $WZ$ production at the LHC including both systematic and statistical effects shows that at  $\sqrt{s}$ = 14 TeV, with 30 ${\rm fb}^{-1}$ luminosity, stronger constraints could be set on a subset of TGCs~\cite{Dobbs:2005ev}. For instance, for the operator inducing an anomalous $WZZ$ coupling, $(1+\Delta g_1^Z) W^{+\mu\nu}W_\mu^-Z_\nu$, the projected 95\% C.L. bound is $-0.0086 < \Delta g_1^Z < 0.011$, compared to the current bound $\Delta g_1^Z = -0.016^{+0.022}_{-0.019}$. A more recent study claimed that the current 7 TeV running with 1 ${\rm fb}^{-1}$ could already improve the present sensitivity to $WZZ$ anomalous couplings at the 2$\sigma$ level~\cite{Eboli:2010qd}. 

Another possibility is that the new physics at the cutoff of the non-linear chiral Lagrangian just generates the required coefficients for the $S$ and $T$ operators with a reference 1 TeV Higgs mass. One can perform a global fit the elecrtroweak precision observables to determine the coefficients of $S$ and $T$ operators at the a cutoff scale $\Lambda$. As shown in Ref.~\cite{Bagger:1999te}, one may learn the required UV physics contributions to $S$ and $T$ as well as the scale for us to anticipate new particles, which could be explored eventually at the 14 TeV LHC.

\section{Conclusions and Discussion}
\label{sec:conclusion}
It is well known that both electroweak precision data and direct Higgs searches disfavor a heavy Higgs with a mass larger than 200 GeV. However, this conclusion is based on the minimal SM, defined as $SU(3)_c \times SU(2)_W \times U(1)_Y$ gauge theory of quarks and leptons with a single elementary Higgs to break the electroweak symmetry. It is likely that new degrees of freedom exist beyond the minimal SM. In principle, these new ingredients, if charged under $SU(2)_W$, could affect the electroweak fit. Furthermore, if they couple to the Higgs, they might also modify the Higgs properties.  In this paper, we explore the possibility of whether including those new degrees of freedom could reconcile a heavy Higgs with the electroweak precision constraints and also evade the current Higgs search. We found that indeed there could be at least two ways to achieve that aim. The first method is to include light color-singlet scalars transforming under $SU(2)_W \times U(1)_Y$, which the Higgs could decay to. If the Higgs mass is below about 300 GeV, the new decay channels could suppress the Higgs partial width to weak gauge bosons by a factor of 2 or more and thus evade the current direct search limit. The second scenario is to include colored scalars transforming under $SU(2)_W \times U(1)_Y$, which could reduce the gluon-fusion production cross section by as much as 90\%. One could test the two possibilities by directly searching for additional particles charged under the weak group. In particular, the colored particles have large production cross sections at the hadron collider. As the decays between different states always yield $W$ bosons, the final states of the long cascades from these new particles could be lepton-rich. Searches involving leptons, such as SS leptons or multi-leptons, will soon be sensitive to a great part of the parameter space of the simplest models we present. Depending on the decays of the lightest state, there could also be model-dependent signals such as multi-jet or multi-photon final states.  Another way to test our proposal is to conduct non-standard Higgs searches. There have already been quite a few studies on non-standard light Higgs decays~\cite{Chang:2006bw, Cheung:2007sva, Chang:2008cw, Carena:2007jk, Falkowski:2010hi}. It would be interesting to check whether these studies directly apply to the heavy Higgs scenario. We also note that we have only explored simplest scenarios with only one additional new particle. It could be useful to consider more complicated possibilities before any definite conclusion about the SM Higgs boson is drawn from the present and future data. 

There are several other directions that warrant further study. Firstly, we focus on the gluon-fusion channel, which has the biggest production cross section at a hadron collider and is the only production channel current LHC searches are sensitive to. There are three other possible Higgs production channels, in particular the vector boson fusion channel has the second largest cross section at the LHC,  only one order of magnitude below that of gluon-fusion at $\sqrt{s}=$ 7 TeV. Its special kinematic features, double forward jets, could also help to drastically suppress various large backgrounds. If the gluon-fusion channel is suppressed by 90\%, the vector-boson fusion channel would become equally important. The 7 TeV run next year will start to be sensitive to this channel and  we expect it to become more important for the 14 TeV run. Secondly, we have only considered single Higgs production in this paper. The deconstructive interference for the single Higgs production from the top quark loop and new colored scalar loop will not guarantee the deconstructive interference for the pair production. On the contrary, the pair production cross section is even constructively enhanced when the deconstructive interference for the single Higgs production is present. This is because the two different operators before EWSB, $[\log{(H^\dagger H)}]G^2$ from the top loop and $H^\dagger H G^2$ from the new colored scalar, have different forms and can not be canceled completely by adjusting the coefficients. From this point of view, it is very important to study double Higgs production and the related signatures at the LHC. Fortunately, at the 14 TeV LHC, Higgs with a 300 GeV mass could be pair produced with a cross section of order fb in the SM. Observation of an enhanced pair production together with suppressed single production would be a concrete confirmation of additional colored scalars.

If there is no sign of a SM Higgs boson at the 7 TeV LHC, another possibility is a very Heavy Higgs boson above the projected search limit at the 7 TeV LHC. In this paper, we have considered several cases including a heavy vector-like fermion mixing with the top quark, a new $Z^\prime$ gauge boson and non-linear realization of EWSB. Although there are no upper limits on the masses of new particles that can be obtained just from EWPT, discovering those new particles would provide us indirect hints about the existence of a very heavy Higgs.

\subsection*{Acknowledgments} 
We would like to thank Nima Arkani-Hamed, Bogdan Dobrescu, Patrick Fox, Michael Peskin, Matt Reece, Neal Weiner  for useful discussions and comments. JiJi Fan thanks the hospitality of SLAC where this work was initiated. Yang Bai thanks the hospitality of IAS where a part of this work was completed. SLAC is operated by Stanford University for the US Department of Energy under contract DE-AC02-76SF00515. JiJi Fan is supported by the DOE grant DE-FG02-91ER40671.

\appendix
\section{Higgs couplings to two gluons and two photons}
\label{sec:HiggsEFT}
Introducing the interaction $-\lambda_2 H^\dagger H {\rm{Tr} [S^2]}$ with $S= S^A T^A$. The effective operator generated from the top quark and color octet loop is~\cite{Boughezal:2010ry} 
\beqa
{\cal L}^{eff} = - C_g \,\frac{h}{v}\,\frac{1}{4} G^a_{\mu\nu}G^{a\mu\nu}  \,,
\eeqa
with
\beq
C_g = - \frac{\alpha_s}{3\,\pi} \left[ \frac{3}{4} A^h_{1/2} (\tau_t) \right] 
 - \frac{\lambda_2\,\alpha_s\,v^2}{8\,\pi\,m_S^2} \left[ 3 A^h_{0} (\tau_S) \right]
\,,
\eeq
and $\tau_t = m^2_h/(4 m_t^2)$ and $\tau_S = m^2_h/(4 m_S^2)$. Here, the form factor functions for fermions, vector gauge bosons and scalars are 
\beqa
A^h_{1/2}(\tau) &=& 2\left[ \tau + (\tau-1) f(\tau)\right] \tau^{-2} \,, \\
A^h_1(\tau) &=& -\left[ 2\tau^2 + 3\tau + 3(2\tau -1) f(\tau)\right] \tau^{-2} \,, \\
A^h_0(\tau) & =& - \left[ \tau - f(\tau) \right] \tau^{-2} \,,
\eeqa
where 
\beqa
f(\tau) = \arcsin^2\sqrt{\tau}\,\quad (\tau\leq 1)\,, \quad \mbox{and}\quad -\frac{1}{4}\left[\log{\frac{1+\sqrt{1-\tau^{-1}}}{1-\sqrt{1-\tau^{-1}}}} -i\pi \right]^2\,\quad (\tau > 1)\,.
\eeqa
When $\tau \rightarrow 0$, $A^h_{1/2} \rightarrow 4/3$, $A^h_{1} \rightarrow -7$ and $A^h_{0} \rightarrow 1/3$.  In the limit $\tau \rightarrow 0$, the NLO results in $\alpha_s^2$ are given by
\beqa
C_g  = - \frac{\alpha_s}{3\,\pi} - \frac{11\,\alpha^2_s}{12\,\pi^2} - \frac{\lambda_2\,\alpha_s\,v^2}{8\,\pi\,m_S^2} - \frac{\lambda_2\,\alpha^2_s\,v^2}{16\,\pi^2\,m_S^2} \left( \frac{33}{2} + 5\,\lambda_{\mathbb{O}} \right)  \,,
\label{eq:NLOcg}
\eeqa
where the coefficient $\lambda_{\mathbb{O}}$ is defined to be the coefficient for the term $-g_s^2 \lambda_{\mathbb{O}} \mbox{Tr}\left[ S^2\right]^2$ in the Lagrangian.

\section{The Functions Used in Section~\ref{sec:QCDfermions} }
\label{sec:Wfunction}
The functions $W_{1}$, $W_{2}$ and $W_{3}$ are defined by
\beqa
W_{1}(y_{1},y_{2}) &\equiv& y_{1}\,+\,y_{2}-\frac{2\,y_{1}\,y_{2}}{y_{1}\,-\,
y_{2}}\log{\frac{y_{1}}{y_{2}}}~,
\nonumber \\ [0.5em]
W_{2}(y_{1},y_{2})&\equiv&\frac{22\,y_{1}\,+\,14\,y_{2}}{9}\,-\,\frac{1}{9}
\log{\frac{y_{1}}{y_{2}}}\,+\,\frac{11\,y_{1}\,+\,1}{18}W_{4}(y_{1},y_{1})\,+\,
\frac{7\,y_{2}\,-\,1}{18}W_{4}(y_{2},y_{2})~,
\nonumber \\ [0.5em]
W_{3}(y_{1},y_{2})&\equiv& \frac{y_{1}\,+\,y_{2}}{2}\,-\,
\frac{(y_{1}-y_{2})^{2}}{3}
\,+\,\left( \frac{(y_{1}-y_{2})^{3}}{6}\,-\,\frac{1}{2}
\frac{y^{2}_{1}+y^{2}_{2}}{y_{1}-y_{2}} \right) \log{\frac{y_{1}}{y_{2}}}\,+\,
\frac{y_{1}-1}{6}W_{4}(y_{1},y_{1}) 
\nonumber \\  
&&\,+\,\frac{y_{2}-1}{6}W_{4}(y_{2},y_{2})
\,+\,\left(  \frac{1}{3}\,-\,\frac{y_{1}+y_{2}}{6}\,-\,\frac{(y_{1}-y_{2})^{2}}{6} 
\right)W_{4}(y_{1},y_{2})~,
\eeqa
with 
\beq
W_{4}(y_{1},y_{2})\equiv \Biggl{\{}
\begin{array}{lll}
 -2\sqrt{\Delta}\,(\arctan{\frac{y_{1}-y_{2}+1}{\sqrt{\Delta}}}-
 \arctan{\frac{y_{1}-y_{2}-1}{\sqrt{\Delta}}})   
 & \Delta\,>\,0  \vspace{3mm}\\
  \sqrt{-\Delta}\,\log{\frac{y_{1}+y_{2}-1+\sqrt{-\Delta}}{y_{1}+y_{2}-1-
  \sqrt{-\Delta}}}  & \Delta\,\le\,0  
\end{array}~,
\eeq
and 
\beq
\Delta\,=\,-1\,-\,y^{2}_{1}\,-\,y^{2}_{2}\,+\,2\,y_{1}\,+\,2\,y_{2}\,+\,2\,
y_{1}\,y_{2}~.
\eeq
Other than $W_{2}$, all $W_{i}(y_{1},y_{2})$ are symmetric functions
under the interchange of the variables $y_{1}$ and $y_{2}$.

\providecommand{\href}[2]{#2}\begingroup\raggedright\endgroup
 \end{document}